\documentclass[lettersize,journal]{IEEEtran}
\usepackage{amsmath,amssymb,amsfonts}
\usepackage{algorithmic}
\usepackage{algorithm}
\usepackage{array}
\usepackage{textcomp}
\usepackage{stfloats}
\usepackage{url}
\usepackage{verbatim}
\usepackage{graphicx}
\usepackage{cite}
\usepackage{bm}
\usepackage{subfigure}
\usepackage{multirow}
\usepackage{threeparttable}
\usepackage{booktabs}
\usepackage{color}

\newtheorem{remark}{Remark}
\newtheorem{assumption}{Assumption}
\newtheorem{property}{Property}
\newtheorem{lemma}{Lemma}
\newtheorem{problem}{Problem}

\newtheorem{definition}{Definition}
\newtheorem{theorem}{Theorem}

\begin{document}

\title{Prescribed-Time Boresight Control of Spacecraft Under \\ Pointing Constraints}

\author{
	\vskip 1em
	Xiaodong Shao,~\IEEEmembership{Member,~IEEE,}
	Haoyang Yang,~\IEEEmembership{Member,~IEEE,}
	Haoran Li,
	Zongyu Zuo,~\IEEEmembership{Senior Member,~IEEE,}
	Jose Guadalupe Romero,~\IEEEmembership{Member,~IEEE,}  
	and Qinglei Hu~\IEEEmembership{Senior Member,~IEEE} 
	
	\thanks{X. Shao, Z. Zuo, and Q. Hu are with the School of Automation Science and Electrical Engineering, Beihang University, Beijing 100191, China (E-mail: xdshao\_sasee@buaa.edu.cn; zzybobby@buaa.edu.cn; huql\_buaa@buaa.edu.cn).}
	\thanks{Haoyang Yang is with the Department of Aeronautical and Aviation Engineering, The Hong Kong Polytechnic University, Hong Kong, SAR 999077, China (E-mail: haoyang.yang@polyu.edu.hk).}
	\thanks{H. Li is with the International Innovation Institute, Beihang University, Hangzhou 311115, China (E-mail: haoranli@buaa.edu.cn).}
	\thanks{J. G. Romero is with the Departamento Académico de Ingeniería Eléctrica y Electrónica, Instituto Tecnológico Autónomo de México (ITAM), Mexico City, Mexico (E-mail: jose.romerovelazquez@itam.mx).}
}

\markboth{Journal of \LaTeX\ Class Files,~Vol.~14, No.~8, August~2021}%
{Shell \MakeLowercase{\textit{et al.}}: A Sample Article Using IEEEtran.cls for IEEE Journals}


\maketitle

\begin{abstract}
	This article proposes an integrated boresight guidance and control (IBGC) scheme to address the boresight reorientation problem of spacecraft under temporal and pointing constraints. A $C^1$ continuous, saturated prescribed-time adjustment (PPTA) function is presented, along with the establishment of a practical prescribed-time stability criterion. Utilizing the time scale transformation technique and the PPTA function, we propose a prescribed-time guidance law that guides the boresight vector from almost any initial orientation in free space to a small neighborhood of the goal orientation within a preassigned time, while avoiding all forbidden zones augmented with safety margins. Subsequently, a prescribed-time disturbance observer (PTDO) is derived to reconstruct the external disturbances. By leveraging barrier and PPTA functions, a PTDO-based reduced-attitude tracking controller is developed, which ensures prescribed-time boresight tracking within a ``safe tube''. By judiciously setting the safety margins, settling times, and safe tube for the guidance and control laws, the proposed IBGC scheme achieves pointing-constrained boresight reorientation within a required task completion time. Simulation and experimental results demonstrate the efficacy of the proposed IBGC scheme.
\end{abstract}

\begin{IEEEkeywords}
	Reduced attitude, prescribed-time control, disturbance observer, pointing constraints.
\end{IEEEkeywords}

\section{Introduction}

\label{sec:introduction}
\IEEEPARstart{R}{eorienting} the boresight axis of light-sensitive payloads (e.g., telescopes, cameras, and star trackers) toward a specified direction is an essential function of many spacecraft. Extensive research has explored three-axis attitude control techniques applicable to spacecraft boresight reorientation \cite{gao2021finite,yang2023optimized,shao2023fault,meng2023second} (just to name a few). Most existing studies formulate the attitude control problem within the framework of Special Orthogonal Group $SO(3)$, or utilizing various attitude parameterizations that can be Euclidean, such as Euler angles and Rodrigues parameters defined in $\mathbb{R}^3$, or non-Euclidean, such as unit quaternions defined on the three-sphere $\mathbb{S}^3$. These attitude representations necessitate knowledge of the spacecraft's full attitude. However, in pointing applications, rotations about the body-fixed boresight axis are usually irrelevant, rendering full attitude information unnecessary. Motivated by this fact, Bullo et al. \cite{bullo1995control} introduced the reduced-attitude representation on the unit 2-sphere $\mathbb{S}^{2}$ to address the boresight reorientation problem. This solution requires only a single boresight vector and three-axis angular velocity measurements. Chaturvedi et al. \cite{chaturvedi2011rigid} proposed a simple proportional-derivative (PD) controller for reduced-attitude stabilization of rigid bodies. Pong and Miller \cite{pong2015reduced} presented reduced-attitude guidance and control laws to achieve pointing, tracking, and searching tasks. 

In pointing applications -- whether for target pointing, inertial stabilization, or interplanetary missions -- spacecraft are typically required to protect onboard light-sensitive payloads from direct exposure to bright objects (e.g., the Sun or the illuminated sides of the Earth) to avoid functional damage \cite{spindler2002attitude}. For example, the James Webb Space Telescope (JWST) must ensure that its infrared instruments avoid direct exposure to the Sun, a critical requirement for maintaining a low-temperature condition necessary for high-precision spectral observations. This requirement imposes pointing constraints on the boresight control, which can be mathematically described as a keep-out cone centered on the direction of the forbidden zone. Existing solutions to the pointing-constrained boresight control problem can generally be classified into two categories: artificial potential function (APF)-based methods \cite{shen2018rigid,yang2021potential,kang2023saturated} and path planning methods \cite{tan2020constrained,celani2020spacecraft,calaon2022constrained}. The APF-based methods, comprising an attractive potential for motion-to-go and a repulsive potential for pointing avoidance, can generate analytical control laws that are both computationally efficient and easy to implement. Hu et al. \cite{hu2019reduced} and Shao et al. \cite{shao2022fault} designed two APF-based reduced-attitude control approaches for spacecraft boresight tracking and reorientation, explicitly accounting for pointing and angular velocity constraints. Recently, Liu et al. \cite{liu2023adaptive} addressed the reduced-attitude boresight control problem under elliptical pointing constraints, employing the navigation function -- a specialized form of APFs. As is well known, the APF-based methods are prone to the local minima problem, which may cause the boresight vector to get trapped in a local minimum, thereby hindering global convergence to the desired orientation. Although the planning methods are free from local minima, they tend to be computationally intensive and demand substantial computing and storage resources.

Nearly all of the aforementioned attitude control approaches achieve only asymptotic convergence of the boresight vector to the desired orientation as time approaches infinity. This is inadequate for time-critical spacecraft reorientation missions, which must be completed within a specified finite time window. Finite/fixed-time control techniques can enable system states to converge to zero within a finite time upper bounded by a computable value \cite{liu2022overview}, offering effective solutions for time-critical boresight control problems. Various finite/fixed-time control approaches have been reported for spacecraft attitude stabilization or tracking, incorporating techniques such as sliding mode control and backstepping control (see, indicatively, \cite{zuo2020robust,xiao2021prescribed,li2023optimal}). In particular, Li et al. \cite{li2023optimal} designed a fixed-time attitude controller to achieve pointing-constrained spacecraft reorientation, by constructing a novel APF-based sliding mode variable. A significant drawback of finite/fixed-time control is that the upper bound on settling time is often overestimated, resulting in unnecessarily large control efforts. Recently, a new finite-time control paradigm, known as prescribed-time control, has been proposed, which allows designers to arbitrarily specify the convergence time in advance. In \cite{munoz2019predefined}, a continuous sliding mode controller was developed to achieve predefined-time stabilization of robotic manipulators. Yucelen et al. \cite{yucelen2018finite} provided an alternative approach for the design of prescribed-time control by introducing a time scale transformation (TST) function. It is noteworthy that TST functions introduce a time-varying gain into the prescribed-time control law, referred to as prescribed-time adjustment (PTA) function \cite{hua2021adaptive}. The PTA function tends to infinity as time approaches the prescribed settling time, which inevitably causes a singularity problem. The lead switching strategy has been widely used to avoid the potential singularity (e.g., see \cite{chen2024time,cheng2023fixed,ning2024dual}). However, this results in a PTA function that is either discontinuous or continuous but not differentiable, limiting its application to the prescribed-time control for high-order systems. These existing prescribed-time control methods, on the other hand, lack the capability to handle pointing constraints.  

To the best of the authors' knowledge, no existing results achieve boresight reorientation of spacecraft under temporal and pointing constraints. To bridge this gap, we propose an integrated boresight guidance and control (IBGC) scheme, using a newly-constructed PTA function. By judiciously setting the safety margins, settling times, and ``safe tube'' for the guidance and control laws, the proposed IBGC scheme ensures that the boresight vector converges to a small neighborhood of the desired orientation within a required task completion time, while avoiding all forbidden pointing zones. The proposed method is validated through simulation and experimental results. The main contributions of this article are three-fold:

\begin{enumerate}
	\item[1)] A $C^1$ continuous, saturated PTA (termed PPTA) function is designed and incorporated into the IBGC framework to achieve prescribed-time boresight control, while a sufficient condition (i.e., Lemma \ref{L_PPTS}) for practical prescribed-time stability is established in the Lyapunov sense. This function, along with Lemma \ref{L_PPTS}, can be readily applied to the design of observers, estimators, and controllers with practical prescribed-time convergence.
	
	\item[2)] An APF-based boresight guidance law is proposed, utilizing the TST technique recently introduced in \cite{yucelen2018finite} and the newly-constructed PPTA function. This guidance law generates a smooth reference trajectory that, starting from almost any initial orientation in free space, converges to a small neighborhood of the desired orientation within a prescribed time, while avoiding all forbidden pointing zones augmented by safety margins. 
	
	\item[3)] A PPTA function-based prescribed-time disturbance observer (PTDO) is developed to reconstruct external disturbances. Then, by leveraging barrier and PPTA functions, a PTDO-based reduced-attitude controller is proposed for boresight tracking. The resulting controller ensures practical prescribed-time convergence of attitude and angular velocity tracking errors, while maintaining the attitude error within a predefined ``safe tube'', even in the presence of external disturbances. 
\end{enumerate}

The remainder of this article is organized as follows. Section \ref{secII} presents a novel PPTA function, outlines the mathematical preliminaries, and formulates the boresight reorientation problem. In Section \ref{secIII}, an IBGC scheme is proposed, accompanied by detailed designs and analyses of prescribed-time boresight guidance and control laws. Sections \ref{secIV} and \ref{secV} provide simulation and experimental results, respectively, to demonstrate the efficacy of the proposed method. Finally, Section \ref{secVI} concludes the article and outlines directions for future work.

\section{Preliminaries and Problem Formulation} \label{secII}

\subsection{Notations}

Throughout the paper, $\bm{I}_{n}$ is the $n\times n$ identity matrix, and $\|\cdot\|$ denotes the $2$-norm of a vector or the induced matrix norm. The special orthogonal group is denoted by $SO(3):=\{\bm{R}\in\mathbb{R}^{3\times3}\mid\bm{R}^{\top}\bm{R}=\bm{I}_{3},~\det(\bm{R})=1\}$. The unit 2-sphere, defined as $\mathbb{S}^{2}:=\{\bm{x}\in\mathbb{R}^{3}\mid\|\bm{x}\|=1\}$, is a 2-dimensional manifold embedded in $\mathbb{R}^{3}$. We denote by $\text{d}_{\mathbb{S}^{2}}(\bm{x},\bm{y}):=\text{arccos}(\bm{x}^{\top}\bm{y})$ the geodesic distance between any two points $\bm{x},\bm{y}\in\mathbb{S}^{2}$. The geodesic distance from a vector $\bm{x}\in\mathbb{S}^{2}$ to a non-empty set $\mathcal{A}$ is defined as $\text{d}_{\mathcal{A}}(\bm{x}):=\inf\{\text{d}_{\mathbb{S}^{2}}(\bm{x},\bm{y})\mid \bm{y}\in\mathcal{A}\}$. The cross product operator $[\cdot]_{\times}:\mathbb{R}^{3} \to \mathbb{R}^{3\times3}$ is defined such that $[\bm{x}]_{\times}\bm{y}=\bm{x}\times\bm{y}$ for any $\bm{x},\bm{y}\in\mathbb{R}^{3}$, which complies with the following properties:	
\begin{align}
&\bm{y}^\top[\bm{y}]_\times=\bm{0},~[\bm{y}]_\times\bm{y}=\bm{0}  \nonumber\\
&[\bm{y}]_\times^\top = -[\bm{y}]_\times,~[\bm{y}]_\times\bm{z}=-[\bm{z}]_\times\bm{y}   \nonumber\\
&[\bm{y}]_\times^2 = \bm{y}\bm{y}^\top-\|\bm{y}\|^2\bm{I}_3,~[\bm{Ry}]_\times = \bm{R}[\bm{y}]_\times\bm{R}^\top \nonumber
\end{align}
for any $\bm{y},\,\bm{z}\in\mathbb{R}^3$ and any $\bm{R}\in SO(3)$. In particular, when $\bm{y}\in\mathbb{S}^2$, it follows that
$$
[\bm{y}]_\times^2 = \bm{y}\bm{y}^\top-\bm{I}_3,~[\bm{y}]_\times^3=-[\bm{y}]_\times.
$$

\subsection{A New PTA Function} \label{secIII-A}

\begin{definition} \label{D_PTS}
	\!\!\!{\rm\cite{zou2022practical}}
	Consider the following system:
	\begin{equation}
	\label{dy}
	\dot{\bm{y}}(t)=\bm{f}(t,\bm{y}(t)),~\bm{f}(t,\bm{0})=0
	\end{equation}
	where $\bm{f}:\mathbb{R}^+\times\mathbb{R}^n\to\mathbb{R}^n$ is a piecewise continuous in $t$ and locally Lipschitz in $\bm{y}$. Let the solution of \eqref{dy} be $\bm{y}(t;\bm{y}_0)$ with $\bm{y}_0:=\bm{y}(0)$. If the origin $\bm{y}=\bm{0}$ of \eqref{dy} is Lyapunov stable, and there exists a constant $T>0$ such that
	\begin{equation}
	\label{PTS}
	\bm{y}(t;\bm{y}_0)=\bm{0},~\forall t\in[T,\infty)
	\end{equation}
	then the origin of \eqref{dy} is prescribed-time stable (PTS). Furthermore, if there exist $\varepsilon>0$ and $T>0$ such that
	\begin{equation}
	\label{PPTS}
	\|\bm{y}(t;\bm{y}_0)\|\leq\varepsilon,~\forall t\in[T,\infty)
	\end{equation} 
	then the origin of \eqref{dy} is termed practical prescribed-time stable (PPTS). If \eqref{PTS} or \eqref{PPTS} hold for all $\bm{y}_0\in\mathbb{R}^n$, then the origin of \eqref{dy} is globally PTS or globally PPTS.
\end{definition} 

\begin{definition} \label{D_TST} 
	A smooth function $\eta:[0,\infty)\to[0,T)$ is called a time scale transformation (TST) function if:
	
	1) it is strictly increasing;
	
	2) it is s.t. $\eta(0)=0$ and $\eta^\prime(0) =1$;
	
	3) it is s.t. $\lim_{s \to \infty}\eta(s)=T$ and $\lim_{s\to \infty}\eta^\prime(s)=0$.
\end{definition}

In fact, a TST function $\eta(s)$ can transform the infinite-time interval $s\in[0,\infty)$ into a finite time interval $t\in[0,T)$. A typical choice of $\eta(s)$ is as follows:
\begin{equation}
\label{eta}
\eta(s)=T(1 - e^{-\frac{s}{T}}).
\end{equation}
Let $\eta^\prime(s):=d\eta(s)/ds$ and define a prescribed-time adjustment (PTA) function  $\mu_p:\mathbb{R}_{\geq0}\to\mathbb{R}_{>0}$ as
\begin{equation}
\label{mu}
\mu_p(t)=\mu_p(\eta(s)):=\dfrac{1}{\eta^\prime(s)}=\dfrac{T}{T-t},~ t\in[0,T)
\end{equation}
which satisfies that $\mu_p(0)=1$ and $\lim_{t\to T^{-}}\mu_p(t)=\infty$. 

It is noteworthy that the prescribed-time controllers utilizing the PTA gain $\mu_p(t)$ cannot be practically implemented due to the unbounded nature of $\mu_p(t)$ as $t\to T$. The commonly-used lead switching approach (e.g., see \cite{chen2024time,cheng2023fixed,ning2024dual}) ensures that the PTA function is well-defined for all $t\geq0$, as described by
\begin{equation}
\nonumber
\mu_s(t)=\left\lbrace
\begin{aligned}
&\mu_p(t),&&\text{if}~~ t\in[0,T^\ast]\\
&1~\text{or}~\mu_p(T^\ast),  &&\text{if}~~t\in[T^\ast,\infty)
\end{aligned}
\right.
\end{equation}
where $T^\ast\in(0,T]$ is a switching time. However, this leads to a discontinuous PTA function, limiting its application in the design of prescribed-time controllers for second- or higher-order systems, since these typically require the PTA function to be at least $C^1$ continuous. To address these issues, we propose a $C^1$ continuous, saturated PTA function, referred to as PPTA function, defined as follows:
\begin{equation}
\label{mu_p}
\mu(t)=	\left\lbrace
\begin{aligned}
&\mu_p(t),&&\text{if}~~ t\in[0,T^\ast]\\
&\mu_p(T^\ast)\psi(t),  &&\text{if}~~t\in[T^\ast,T) \\	&\dfrac{\pi+2}{\pi}\mu_p(T^\ast),&&\text{if}~~ t\in[T,\infty)
\end{aligned}
\right.
\end{equation}
with 
\begin{equation}
\nonumber
\psi(t):=\left[1 + \dfrac{2}{\pi} \sin\left(\dfrac{\pi}{2}\dfrac{t - T^\ast}{T - T^\ast}\right) \right].
\end{equation}

\begin{property} \label{Pro_PTA}
	The newly-constructed PPTA function $\mu(t)$ satisfies the following properties:	
	\begin{enumerate}
		\item it is $C^1$ continuous and bounded on $t\in[0,\infty)$;
		\item it is monotonically increasing on $t\in[0,T]$;
		\item its initial value satisfies $\mu(0)=1$;
		\item for all $t\geq T$, $\mu(t)=\left( 1+\frac{2}{\pi}\right)\frac{T}{T-T^\ast}$. 
	\end{enumerate}
\end{property}

\begin{figure}[hbt!]
	\centering
	\includegraphics[width=8.5cm]{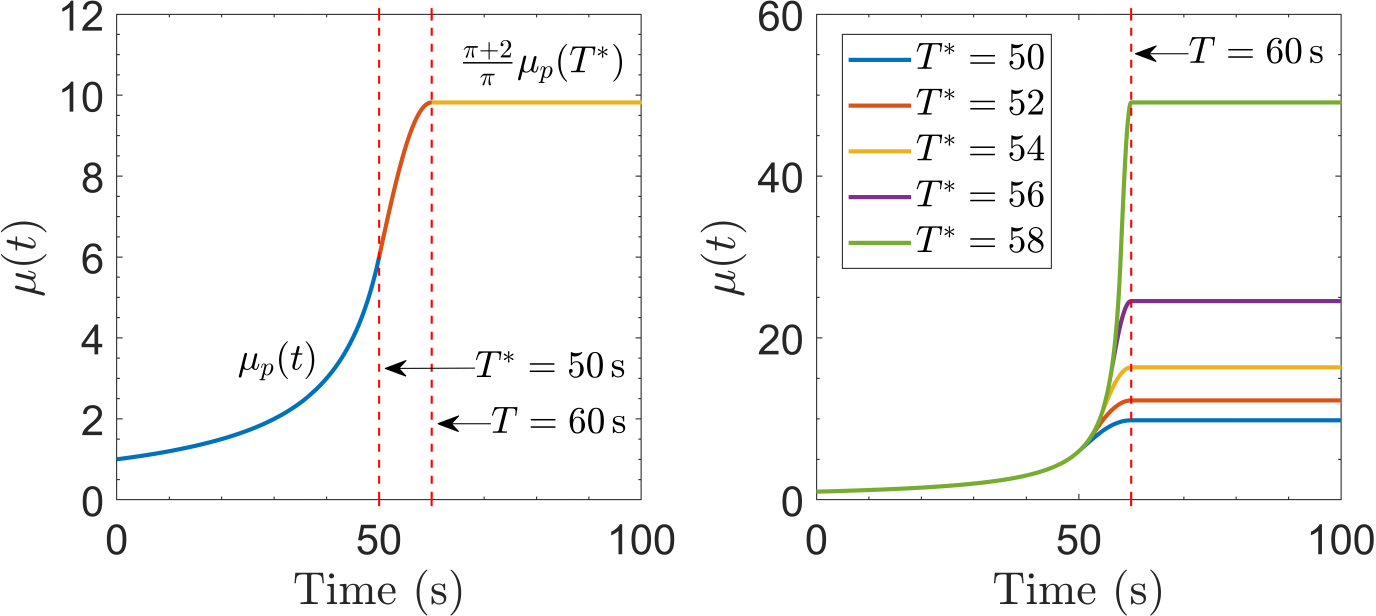}
	\caption{Time history of $\mu(t)$: (a) the left subfigure illustrates a special case of $\mu(t)$ with $T^\ast=50$ and $T=60$; (b) the right subfigure depicts the time histories of $\mu(t)$ for different values of $T^\ast$.}
	\label{simu_mu_p}
\end{figure}

The properties of $\mu(t)$ described above are straightforward and can be clearly seen in Fig. \ref{simu_mu_p}. In the following, we establish a sufficient condition for PPTS of the system's origin.

\begin{lemma}\label{L_PPTS}
	Consider the nonlinear system described by \eqref{dy}. Suppose there are $C^1$ continuous, positive definite function $V(\bm{y})$, defined on a neighborhood $\mathbb{U}\subset \mathbb{R}^n$ of the origin, and real numbers $\alpha>1/T$ and $\beta\geq0$, such that 
	\begin{align}
	\label{dV_PTS}
	\dot{V}(\bm{y})\leq - \alpha \mu(t)V(\bm{y}) + \beta,~\forall \bm{y}\in \mathbb{U}
	\end{align}
	where $\mu(t)$ is a PPTA function defined in \eqref{mu_p}. Then, the origin of \eqref{dy} is locally PPTS, and the solution of \eqref{dy} converges to the following set within the prescribed time $T^\ast$:
	\begin{equation}
	\label{Omega}
	\Omega:=\left\lbrace \bm{y}\in\mathbb{R}^n \mid V(\bm{y})\leq \max\left\lbrace v_1, v_2\right\rbrace \right\rbrace 
	\end{equation}
	where $v_1$ and $v_2$ are given by
	\begin{align}
	\label{V_e}
	v_1 := \left(\dfrac{T-T^\ast}{T}\right) ^{\alpha T}V_0+\bar{v}_1,~v_2 := \dfrac{\beta(T-T^\ast)}{\alpha T}
	\end{align}
	with $V_0:=V(\bm{y}_0)$ and
	\begin{equation}
	\nonumber
	\bar{v}_1:=\dfrac{\beta T}{\alpha T-1}\left[\dfrac{T-T^\ast}{T} - \left(\dfrac{T-T^\ast}{T}\right)^{\alpha T} \right]>0.
	\end{equation}
	By choosing $\alpha$ and $T^\ast$ such that $T-T^\ast\ll \alpha T-1$, the set $\Omega$ can be arbitrarily reduced by increasing $T^\ast$ closer to $T$. If $\mathbb{U}=\mathbb{R}^n$ and $V(\bm{y})$ is radially unbounded, then the origin of \eqref{dy} is globally PPTS.
\end{lemma}

\textit{Proof}. The proof is relegated to Appendix A. $\hfill \blacksquare$	

\begin{remark} \label{R1}	
	The PPTA function $\mu(t)$, along with Lemma \ref{L_PPTS}, can be used to design prescribed-time state/disturbance observers, parameter estimators, and tracking controllers. Note that designing prescribed-time controllers for $n$-order systems typically requires the PPTA function to be $C^{n-1}$ continuous. This limits the direct application of $\mu(t)$ to high-order systems, as it is only $C^1$ continuous. In such cases, the dynamic surface control approach presented in \cite{swaroop2000dynamic} can be used in conjunction with backstepping control. Additionally, while a higher value of $\mu(t)$ on $t\in[T^\ast,\infty)$ is instrumental for reducing the steady-state error bound, this may excite unmodeled high-frequency dynamics in practice, potentially leading to instability of the closed system. Therefore, for a prescribed time $T$, one should select an appropriate $T^\ast$ to strike a balance between control accuracy and system stability. 
\end{remark}


\subsection{Problem Formulation}

Denote by $\bm{R}\in SO(3)$ and $\bm{\omega}\in\mathbb{R}^{3}$ the attitude and angular velocity of the body-fixed frame $\mathcal{B}$ of a spacecraft with respect to (w.r.t.) an inertial frame $\mathcal{N}$, expressed in $\mathcal{B}$. The spacecraft attitude dynamics are described by 
\begin{align}
\dot{\bm{R}}&=\bm{R}[\bm{\omega}]_{\times} \label{dR} \\
\bm{J}\dot{\bm{\omega}}&=[\bm{J}\bm{\omega}]_{\times}\bm{\omega}+\bm{u}+\bm{d} \label{domega}
\end{align}
where $\bm{J}=\bm{J}^\top\in\mathbb{R}^{3 \times 3}$ is the positive-definite inertia matrix of spacecraft, while $\bm{u}\in\mathbb{R}^{3}$ and $\bm{d}\in\mathbb{R}^{3}$ denote the control and disturbance torques, respectively. Let $\bm{b}\in\mathbb{S}^{2}$ be the boresight vector of an onboard sensitive payload (e.g., telescope, camera or other instruments) resolved in the frame $\mathcal{B}$. We disregard the rotation about this axis and express it in the frame $\mathcal{N}$ as $\bm{x}:=\bm{R}\bm{b}\in\mathbb{S}^{2}$. The kinematics of $\bm{x}$ is expressed as
\begin{align}
\label{dx}
\dot{\bm{x}}=[\bm{\Omega}]_{\times}\bm{x}
\end{align}  
where $\bm{\Omega}:=\bm{R}\bm{\omega}$ is the angular velocity expressed in $\mathcal{N}$. For any $(\bm{x},\bm{\Omega})\in\mathbb{S}^2\times \mathbb{R}^3$, $[\bm{\Omega}]_{\times}\bm{x} \in \mathbf{T}_{\bm{x}}\mathbb{S}^{2}$, indicating that $\bm{x}\in\mathbb{S}^{2}$ always holds under the driftless kinematics \eqref{dx}. 

\begin{assumption} \label{A1}
	The disturbance torque $\bm{d}$ is differentiable and bounded, and there exists an unknown constant $d_m>0$ such that $\|\dot{\bm{d}}\|\leq d_m$.
\end{assumption}

\begin{figure}[hbt!]
	\centering
	\includegraphics[width=6cm]{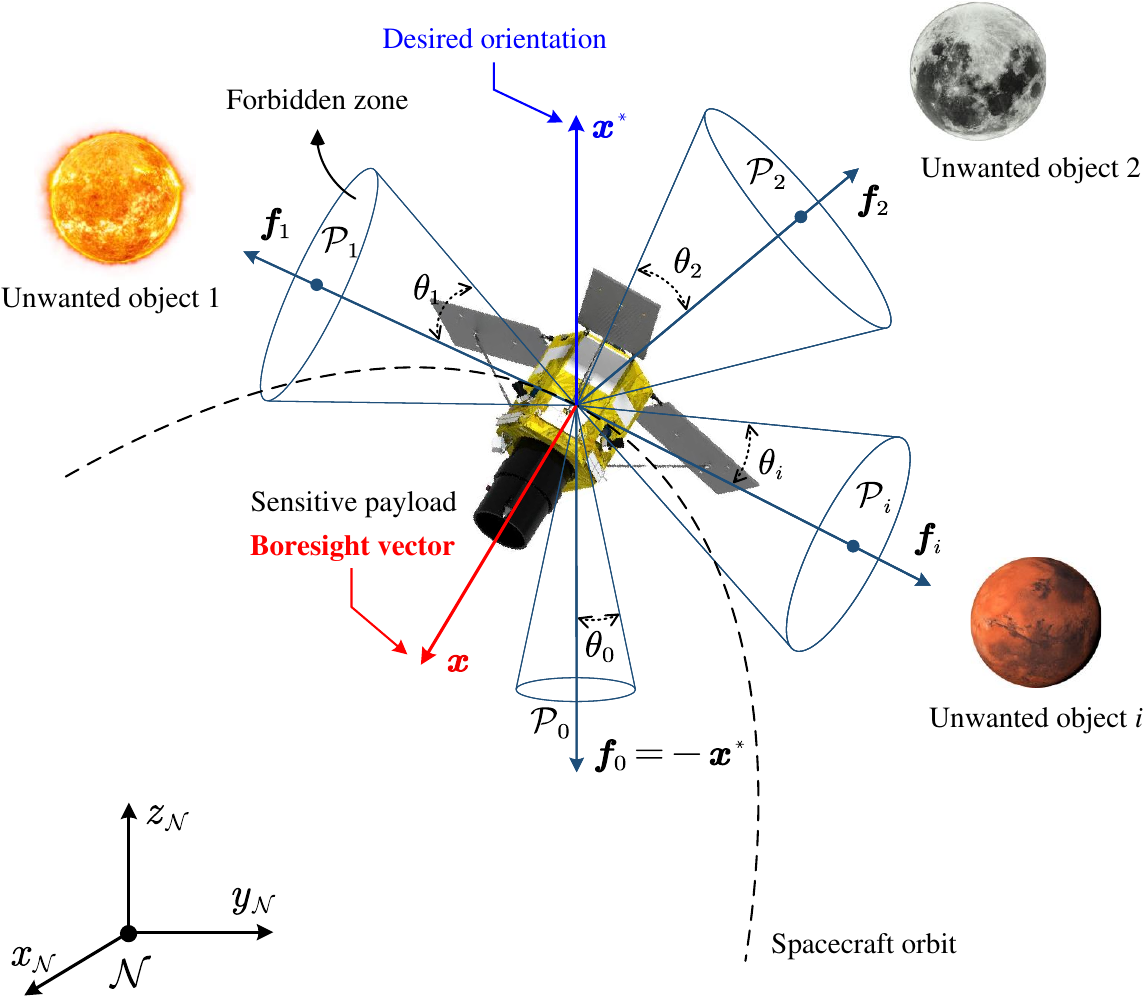}
	\caption{Illustration of boresight pointing constraints.}
	\label{pointing_constraint}
\end{figure} 

During the boresight reorientation, it is necessary to ensure that the boresight axis of the on-board sensitive payload avoids bright objects, such as the Sun or the illuminated sides of the Earth and Moon. Suppose that there are $m\geq1$ bright objects that the sensitive payload should avoid, resulting in $m$ cone-shaped forbidden pointing zones, as illustrated in Fig. \ref{pointing_constraint}. The forbidden pointing constraints can be described in the inertial frame $\mathcal{N}$ as follows:
\begin{equation}
\nonumber
\mathcal{P}_{i}:=\{\bm{x}\in\mathbb{S}^{2}\mid\text{d}_{\mathbb{S}^{2}}(\bm{x},\bm{f}_{i}) \leq \theta_i\},~i\in\mathbb{I}=\{0,1,...,m\}
\end{equation}
where $\bm{f}_{i}\in\mathbb{S}^{2}$ and $\theta_{i}\in(0,\pi/2)$ denote the center axis and half-angle of the $i$-th forbidden pointing cone, respectively. Note that $\mathcal{P}_{0}$ is a virtual forbidden zone around the antipodal point of the desired orientation $\bm{x}^\ast$ (i.e., $\bm{f}_0=-\bm{x}^\ast$). This zone is introduced to prevent $\bm{x}$ from converging to the undesired equilibrium $-\bm{x}^\ast$ under a continuous controller. Then, the free space of the boresight vector $\bm{x}$ is
\begin{equation}
\nonumber
\mathcal{F}:=\mathbb{S}^{2} \setminus \bigcup  \limits_{i=0}^{m} \mathcal{P}_{i}=\{\bm{x}\in\mathbb{S}^{2}\mid\text{d}_{\mathbb{S}^{2}}(\bm{x},\bm{f}_{i})>\theta_{i},~\forall i\in\mathbb{I}\}.
\end{equation}

To ensure that the boresight vector $\bm{x}$ can fit freely between any of the forbidden zones, we make the following ``isolated'' constraint assumption:

\begin{assumption} \label{A2}
	The forbidden pointing zones are separated from each other by a clearance of at least
	\begin{equation}
	\nonumber
	{\rm{d}}_{\mathbb{S}^{2}}(\bm{f}_{i},\bm{f}_{j})>\theta_{i}+\theta_{j}+2\iota,~\forall i,j\in \mathbb{I},~i \neq j
	\end{equation}
	where $\iota\in(0,\pi/2)$ is a constant.  
\end{assumption}

The problem to be addressed is summarized below:

\begin{problem} \label{Pr1}
	Consider the spacecraft attitude dynamics given by \eqref{dR} and \eqref{domega} under Assumption \ref{A1}, and suppose that the forbidden zones $\mathcal{P}_i$, $i\in\mathbb{I}$ satisfy Assumption \ref{A2}. The control objective is to develop a controller $\bm{u}$ that guides the boresight vector $\bm{x}$ from an initial pointing $\bm{x}(0)\in \mathcal{F}$ to the goal pointing $\bm{x}^{\ast}\in\mathcal{F}$ (motion to goal) within a designer-specified time $T$, while ensuring that $\bm{x}$ remains within the free space $\mathcal{F}$ at all times (forbidden pointing avoidance), despite the presence of inertia uncertainties.
\end{problem}

\section{Main Results} \label{secIII}

\begin{figure}[hbt!]
	\centering
	\includegraphics[width=8.5cm]{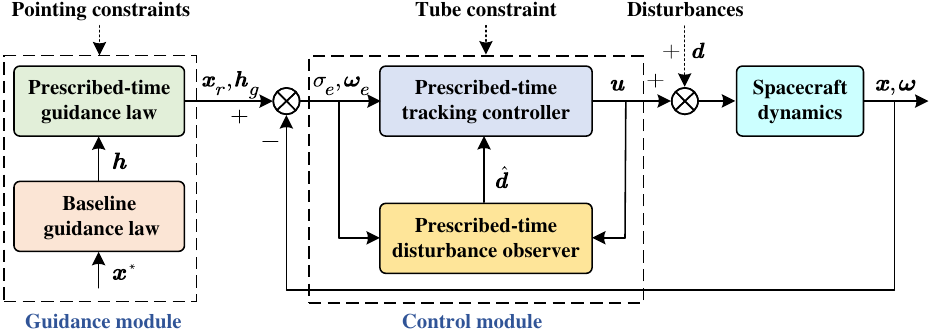}
	\caption{Block diagram of the proposed IBGC scheme.}
	\label{control_framework}
\end{figure}

In this section, an IBGC scheme that incorporates the PPTA function, defined in \eqref{mu_p}, is proposed to address Problem \ref{Pr1}. The block diagram of this scheme is shown in Fig.\ref{control_framework}, where the guidance module accounts for generating a smooth reference trajectory $\bm{x}_r$ with pointing avoidance and practical prescribed-time convergence. While in the control module, a PTDO-based reduced-attitude controller is developed to achieve prescribed-time boresight tracking within a predefined ``safe tube'', with a radius no larger than the safety margin, even in the presence of external disturbances. In this context, the boresight vector $\bm{x}$ will converge to $\bm{x}^\ast$ within a required task completion time, while avoiding all forbidden zones. 

\subsection{Prescribed-Time Boresight Guidance}

Before proceeding, let $\bm{x}_r\in\mathbb{S}^2$ and $\bm{\Omega}_r\in\mathbb{R}^3$ denote, respectively, the reference boresight vector and angular velocity resolved in the inertial frame. As per \eqref{dx}, the kinematics of $\bm{x}_r$ is expressed as
\begin{align}
\label{dx_r}
\dot{\bm{x}}_r=[\bm{\Omega}_r]_{\times}\bm{x}_r
\end{align} 
with $\bm{x}_r(0)=\bm{x}(0)$ and $\bm{\Omega}_r(0)=\bm{\Omega}(0)$. 

To enhance safety, a safety margin of size $\epsilon<\iota$ is defined around each forbidden zone $\mathcal{P}_{i}$, leading to $m+1$ augmented forbidden zones
\begin{equation}
\nonumber
\mathcal{P}_{i}^{\epsilon}=\{\bm{x}\in\mathbb{S}^{2}\mid\text{d}_{\mathbb{S}^{2}}(\bm{x},\bm{f}_{i}) \leq \theta_i+\epsilon\},~i\in\mathbb{I}
\end{equation}
and an eroded free space $\mathcal{F}_{\epsilon}:=\mathbb{S}^2 \setminus \bigcup \nolimits_{i=0}^{m}\mathcal{P}_{i}^{\epsilon}$. Furthermore, an influence area is introduced for $\mathcal{P}_{i}$, defined as $\mathcal{I}_i:=\{\bm{x}\in\mathbb{S}^{2}\mid\text{d}_{\mathcal{P}_i}(\bm{x}) \leq \epsilon^\ast\}$ with
\begin{equation}
\label{epsilon_cond}
0<\epsilon<\epsilon^{\ast}\leq\iota.
\end{equation}
Bearing Assumption \ref{A2} and \eqref{epsilon_cond} in mind, the influence regions of different pointing constraints do not overlap. Thus, there is at most one forbidden pointing zone that can activate pointing avoidance at any given time. Note that $\epsilon^\ast$ should be chosen such that $\bm{x}^\ast$ lies outside the influence regions of all forbidden pointing constraints $\mathcal{P}_i$, $i\in\mathbb{I}$. As a stepping stone, we propose a baseline boresight guidance law that asymptotically guides $\bm{x}_r$ toward the goal $\bm{x}^{\ast}$ (motion to goal) from almost all initial orientations $\bm{x}_r(0)\in\mathcal{F}_{\epsilon}$, while ensuring forward invariance of the free space $\mathcal{F}_{\epsilon}$ (forbidden pointing avoidance). 

Consider the following APF $U:\mathcal{F}_{\varepsilon}\to\mathbb{R}^{+}$
\begin{equation}
\label{U}
U(\bm{x}_r) = \underbrace{k_a(1-\bm{x}_r^\top\bm{x}^\ast)}_{\text{attractive potential}} + \underbrace{k_r \sum_{i=0}^{m}\phi_i(\bm{x}_r^\top\bm{f}_i)}_{\text{repulsive potential}}
\end{equation}
where $k_a,\,k_r>0$ are constants, and $\phi_i$ is a smooth repulsive function given by
\begin{equation}
\nonumber
\phi_i(z)=\left\lbrace
\begin{aligned}
&(z-\varepsilon_i^\ast)^2\text{ln}\frac{\varepsilon_i-\varepsilon_i^\ast}{\varepsilon_i-z},&&\text{if}~ z\in[\varepsilon_i^\ast,\varepsilon_i)\\
&0,&&\text{if}~z\leq\varepsilon_i^\ast
\end{aligned}
\right.
\end{equation}
with its gradient given by
\begin{equation}
\nonumber
\nabla\phi_i(z)=\left\lbrace
\begin{aligned}
&2(z-\varepsilon_i^\ast)\text{ln}\frac{\varepsilon_i-\varepsilon_i^\ast}{\varepsilon_i-z}+\frac{(z-\varepsilon_i^\ast)^2}{\varepsilon_i-z},&&\text{if}~ z\in[\varepsilon_i^\ast,\varepsilon_i)\\
&0,&&\text{if}~z\leq\varepsilon_i^\ast
\end{aligned}
\right.
\end{equation}
where $\varepsilon_i:=\cos(\theta_i+\epsilon)$ and $\varepsilon_i^\ast:=\cos(\theta_i+\epsilon^\ast)$. It is easy to verify that: 1) $\phi_i(z)$ is continuously differentiable; 2) $\phi_i(z)$ is strictly increasing on $z\in[\varepsilon_i^\ast,\varepsilon_i)$, with $\lim_{z\to\varepsilon_i}\phi_i(z)=+\infty$ and $\phi_i(z)=0$ for all $z\leq\varepsilon_i^\ast$; 3)  $\nabla\phi_i(z)$ is continuously differentiable and $\nabla\phi_i(z)\geq0$ for all $z<\varepsilon_i$. As $\phi_i(\bm{x}_r^\top\bm{f}_i)\geq0$ for all $\bm{x}_r^\top\bm{f}_i<\varepsilon_i$, the APF $U(\bm{x}_r)$ is positive definite on the practical free space $\mathcal{F}_\epsilon$, and the equilibrium $\bm{x}_r=\bm{x}^\ast$ is the unique global minimum of $U(\bm{x}_r)$. The gradient of $U(\bm{x}_r)$ is derived as
\begin{equation}
\label{nabla_U}
\nabla U(\bm{x}_r)=-k_a\bm{x}^{\ast}+k_r\sum_{i=0}^{m}\left(\nabla\phi_i(\bm{x}_r^\top\bm{f}_i)\cdot\bm{f}_i\right).
\end{equation}

Design the baseline guidance law as 
\begin{equation}
\label{h}
\bm{\Omega}_r=\bm{h}(\bm{x}_r):=[\bm{x}_r]_\times^\top\nabla U(\bm{x}_r).
\end{equation}
Substituting \eqref{h} into \eqref{dx} yields 
\begin{equation}
\label{dx_h}
\dot{\bm{x}}_r=-[\bm{x}_r]_\times \bm{h}(\bm{x}_r) =-[\bm{x}_r]_\times[\bm{x}_r]_\times^\top \nabla U(\bm{x}_r).
\end{equation}

\begin{theorem} \label{L_baseline}
	Consider the boresight kinematics described by \eqref{dx}. Suppose that the forbidden constraints $\mathcal{P}_i$, $i\in\mathbb{I}$ satisfy Assumption \ref{A2} and that their corresponding safety margin and influence region comply with \eqref{epsilon_cond}. Then, the baseline guidance law $\bm{\Omega}_r=\bm{h}(\bm{x}_r)$ given by \eqref{h} guarantees that:
	\begin{enumerate}
		\item The practical free space $\mathcal{F}_{\varepsilon}$ is forward invariant.
		\item The boresight vector $\bm{x}_r$ asymptotically converges to the desired equilibrium $\bm{x}^\ast$ or reaches a set of $m$ undesired critical points $\bigcup_{i=1}^{m}\{\bm{c}_i\}$ with $\bm{c}_i$ satisfying 
		\begin{equation}
		\label{Crit}
		k_a[\bm{c}_i]_\times\bm{x}^{\ast}=k_r\nabla\phi_i(\bm{c}_i^\top\bm{f}_i)[\bm{c}_i]_\times\bm{f}_i.
		\end{equation}
	\end{enumerate} 
\end{theorem}

\textit{Proof}. The proof is relegated to Appendix B. $\hfill \blacksquare$	

To achieve prescribed-time boresight guidance, inspired by \cite{shakouri2021prescribed}, we propose a prescribed-time guidance law by incorporating a PPTA function $\mu_g(t)$ with design parameters $T_g$ and $T_{g}^\ast$ into the baseline law \eqref{h}
\begin{equation}
\label{h_g}
\bm{\Omega}_r=\bm{h}_g(\bm{x}_r,t):=\mu_g(t)\bm{h}(\bm{x}_r).
\end{equation} 
As $\bm{h}(\bm{x}_r)$ is smooth and $\mu_g(t)$ is $C^1$ continuous on $t\in[0,\infty)$, $\bm{h}_g(\bm{x}_r,t)$ is continuously differentiable for all $t\geq0$. Now, the kinematic equation \eqref{dx} is rewritten as
\begin{equation}
\label{dx_mu_h}
\dot{\bm{x}}_r=-\mu_g(t)[\bm{x}_r]_\times\bm{h}(\bm{x}_r).
\end{equation}

\begin{theorem} \label{Th1}
	Consider the boresight kinematics described by \eqref{dx}. The extended guidance law $\bm{\Omega}_r=\bm{h}_g(\bm{x}_r,t)$, as defined in \eqref{h_g}, ensures that the boresight vector $\bm{x}_r$ converges to a small neighborhood around either the goal point $\bm{x}^\ast$ or an undesired critical point $\bm{c}_i$ within the prescribed time $T_{g}^\ast$. 
\end{theorem}

\textit{Proof}. The proof is analyzed over two time intervals $[0,T_{g}^\ast)$ and $[T_{g}^\ast,\infty)$, separately.

1) Consider the interval $t\in[0,T_{g}^\ast)$. Let us define $t:=\eta(s)$ (a TST function in Definition \ref{D_TST}), $\bar{\bm{x}}_r(s):=\bm{x}_r(t)=\bm{x}_r(\eta(s))$. Then, the kinematics \eqref{dx_mu_h} can be  rewritten in a stretched time interval $s\in[0,s_g)$ as 
\begin{equation}
\label{dx_pri}
\bar{\bm{x}}_r^\prime(s):=\dfrac{d\bar{\bm{x}}_r(s)}{ds}=\dfrac{d\bm{x}_r(\eta(s))}{d\eta(s)}\cdot\dfrac{d\eta(s)}{ds}
\end{equation}
where $s_g:=\eta^{-1}(T_{g}^\ast)=-T_g\ln(1-T_{g}^\ast/T_g)$ is a constant. It is evident that $s_g\to+\infty$ as $T_{g}^\ast\to T_g$. With \eqref{mu} and \eqref{dx_mu_h} in mind, \eqref{dx_pri} reduces to
\begin{align}
\label{dx_pri_bar}
\bar{\bm{x}}_r^\prime(s)&=-\mu_g(\eta(s))[\bm{x}_r(\eta(s))]_\times\bm{h}(\bm{x}_r(\eta(s)))\eta^\prime(s) \nonumber \\
&=-[\bar{\bm{x}}_r(s)]_\times\bm{h}(\bar{\bm{x}}_r(s))
\end{align}
where the facts that $\mu_g(t)=\mu_p(t)$ for all $t\in[0,T_{g}^\ast)$ and that $\mu_p(\eta(s))\eta^\prime(s)=1$ have been used. As per Definition \ref{D_TST}, we have $\eta(0)=0$ and $\mu_p(0)=1$, from which it follows that $\bar{\bm{x}}_r(0)=\bm{x}_r(0)$ and $\bar{\bm{x}}_r^\prime(0)=\dot{\bm{x}}_r(0)$. Therefore, the solution of \eqref{dx_pri_bar} is equivalent to that of the kinematics \eqref{dx_h}. In Theorem \ref{L_baseline}, we have demonstrated that the solution of \eqref{dx_h} asymptotically converges to the set $\{\bm{x}^\ast\}\bigcup_{i=1}^{m}\{\bm{c}_i\}$, indicating that $\bar{\bm{x}}_r(s)$ will converge to a neighborhood of either the desired equilibrium $\bm{x}^\ast$ or an undesired critical point $\bm{c}_i$ ($i\in\mathbb{I}\setminus\{0\}$) within the stretched time interval $s\in[0,s_g)$. Moreover, as $s_g$ increases (by setting $T_{g}^\ast$ closer to $T_g$), the neighborhood size decreases. Note that $\bm{x}_r(t)=\bar{\bm{x}}_r(s)$ and $t\to T_{g}^\ast$ as $s\to s_g$. Then, it can be claimed that $\bm{x}_r(t)$ converges to a neighborhood of either $\bm{x}^\ast$ or $\bm{c}_i$ within the prescribed time $T_{g}^\ast$. Furthermore, inspecting Theorem \ref{L_baseline} reveals that $\bar{\bm{x}}_r(s)\in\mathcal{F}_\epsilon$ for all $s\in[0,s_g)$, allowing us to conclude that $\bm{x}_r(t)$ remains within the practical free space $\mathcal{F}_\epsilon$ for all $t\in[0,T_{g}^\ast)$. Actually, the solution of \eqref{dx_mu_h} on the interval $t\in[0,T_{g}^\ast)$ follows the same path as that obtained by the baseline guidance law on $s\in[0,s_g)$. 

2) Next, consider the interval $t\in[T_{g}^\ast,\infty)$. Since $\mu_g(t)\geq\mu_p(T_{g}^\ast)>1$, $\forall t\geq T_{g}^\ast$ holds due to Property \ref{Pro_PTA}, the trajectory $\bm{x}_r(t)$, governed by \eqref{dx_mu_h}, will continuously and asymptotically converge to either the desired equilibrium $\bm{x}^\ast$ or the undesired critical point $\bm{c}_i$, while avoiding all forbidden pointing zones. This is evident from the proof of Theorem \ref{L_baseline}.


Synthesizing the above analyses allow us to conclude Theorem \ref{Th1}. This completes the proof.  $\hfill \blacksquare$ 

\subsection{Prescribed-Time Boresight Tracking Control}

By leveraging barrier and PPTA functions, we here propose a PTDO-based reduced-attitude tracking controller to achieve reference trajectory tracking within a prescribed time $T_c^\ast\leq T_g^\ast$, while ensuring that the boresight vector $\bm{x}$ remains within a fixed-size ``safe tube'' around the reference trajectory $\bm{x}_r$. To facilitate the subsequent analysis, the reference trajectory $\bm{x}_r\in\mathbb{S}^2$, governed by \eqref{dx_r} with $\bm{\Omega}_r$ defined in \eqref{h_g}, is expressed in the body frame as 
\begin{equation}
\nonumber
\bm{\sigma}: = \bm{R}^\top\bm{x}_r \in\mathbb{S}^2
\end{equation}
which is the \textit{reduced attitude}. 

The objective of boresight tracking is to align the boresight vector $\bm{x}$ with the reference trajectory $\bm{x}_r$ in the inertial frame. This is equivalent to stabilizing $\bm{\sigma}$ to the body-fixed boresight vector $\bm{b}$ in the body frame. In the following, we formulate a tracking problem for reduced attitude and define the reduced-attitude tracking error in the body frame as 
\begin{align}
\label{err}
\sigma_e &= 1 - (\bm{x}_r^\top\bm{R})(\bm{R}^\top\bm{x}) \nonumber \\
&= 1 - \bm{\sigma}^\top\bm{b}.
\end{align}
Given the kinematics of \eqref{dR} and \eqref{dx_r}, we apply the properties of cross product operator to the time derivative of $\sigma_e$, yielding the reduced-attitude tracking error kinematics
\begin{align}
\label{dx_e}
\dot{\sigma}_e &= -\bm{b}^\top(\dot{\bm{R}}^\top\bm{x}_r+\bm{R}^\top\dot{\bm{x}}_r)   \nonumber \\
&=-\bm{b}^\top\left((\bm{R}[\bm{\omega}]_\times)^\top\bm{x}_r+\bm{R}^\top[\bm{h}_p]_\times \bm{R}\bm{R}^\top\bm{x}_r\right) \nonumber \\
&=-\bm{b}^\top[\bm{\sigma}]_\times \bm{\omega}_e
\end{align}
where $\bm{\omega}_e\in\mathbb{R}^3$ denotes the angular velocity error expressed in the body frame, defined as
\begin{equation}
\label{omega_e}
\bm{\omega}_e:=\bm{\omega}-\bm{R}^\top\bm{h}_p.
\end{equation}
Then, applying \eqref{omega_e} gets the error dynamics
\begin{equation}
\label{Jd_omega}
\bm{J}\dot{\bm{\omega}}_e = \underbrace{-\bm{C}\bm{\omega}_e - \bm{G}}_{\bm{H}} + \bm{u} + \bm{d}
\end{equation}
with 
\begin{align}
\bm{C}&:=-[\bm{J}(\bm{\omega}_e + \bm{R}^\top\bm{h}_p)]_\times+([\bm{R}^\top\bm{h}_p]_\times\bm{J} + \bm{J}[\bm{R}^\top\bm{h}_p]_\times) \nonumber \\
\bm{G}&:=[\bm{R}^\top\bm{h}_p]_\times\bm{J}\bm{R}^\top\bm{h}_p + \bm{J}\bm{R}^\top\dot{\bm{h}}_p. \nonumber
\end{align}
Here, the augments of $\bm{C}$ and $\bm{G}$ are omitted for brevity. 

A PTDO is designed to reconstruct the disturbance $\bm{d}$:
\begin{align}
\label{dp}
\dot{\bm{p}}= & - c_1 \mu_c(t) \bm{p} - c_1 \mu_c(t)(c_1 \mu_c(t)\bm{J}\bm{\omega}_e + \bm{H} + \bm{u}) \nonumber \\
&- c_1 \dot{\mu}_c(t)\bm{J}\bm{\omega}_e 
\end{align}
\begin{equation}
\label{hat_d}
\hat{\bm{d}}= \bm{p} + c_1 \mu_c(t)\bm{J}\bm{\omega}_e
\end{equation}
where $c_1>0$ is a constant; $\mu_c(t)$ is a PPTA function defined in \eqref{mu_p} with design parameters $T_c$ and $T_{c}^\ast$ ($T_{c}^\ast\leq T_{g}^\ast$); $\bm{p}\in\mathbb{R}^3$ is the internal state of the observer, and $\hat{\bm{d}}\in\mathbb{R}^3$ is the estimated disturbance. Define the estimation error as $\tilde{\bm{d}}:=\bm{d}-\hat{\bm{d}}$. Using \eqref{dp} and \eqref{hat_d}, the observation error dynamics is given by
\begin{align}
\label{dot_d_tilde}
\dot{\tilde{\bm{d}}} & = \dot{\bm{d}} - (\dot{\bm{p}} + c_1 \mu_c(t)\bm{J}\dot{\bm{\omega}}_e + c_1 \dot{\mu}_c(t)\bm{J}\bm{\omega}_e)  \nonumber \\
& = \dot{\bm{d}} - c_1 \mu_c(t) \tilde{\bm{d}}.
\end{align}

\begin{theorem}
	Consider the reduced-attitude error dynamics described in \eqref{Jd_omega}. The PTDO given by \eqref{dp} and \eqref{hat_d} is globally PPTS in the sense that the estimation error $\tilde{\bm{d}}$ remains within the set $\|\tilde{\bm{d}}\|\leq \tilde{d}_m :=\max\{ d_m/c_1,\|\tilde{\bm{d}}(0)\|\}$ and converges to a small neighborhood of zero within the prescribed time $T_{c}^\ast$.
\end{theorem}

\textit{Proof}. Consider the Lyapunov function candidate
\begin{equation}
V_d = \dfrac{1}{2}\tilde{\bm{d}}^\top\tilde{\bm{d}}
\end{equation}
In view of \eqref{dot_d_tilde}, the time derivative of $V_d$ is
\begin{equation}
\label{dV_d}
\dot{V}_d = - c_1 \mu_c(t) \tilde{\bm{d}}^\top\tilde{\bm{d}} + \tilde{\bm{d}}^\top \dot{\bm{d}}. 
\end{equation}
Using Young's inequality and Assumption \ref{A1}, we have
\begin{equation}
\label{Young}
\tilde{\bm{d}}^\top \dot{\bm{d}}\leq \dfrac{c_1 \mu_c(t)}{2}\|\tilde{\bm{d}}\|^2 + \dfrac{1}{2c_1} d_m^2
\end{equation}
wherein we have used the fact that $\mu_c(t)\geq \mu_c(0)=1$. Then, inserting \eqref{Young} into \eqref{dV_d} yields
\begin{align}
\label{dV_d1}
\dot{V}_d \leq - c_1 V_d + \dfrac{1}{2c_1} d_m^2
\end{align}
from which it is clear that $\dot{V}_d\leq 0$ if $\|\tilde{\bm{d}}\|\geq d_m/c_1$, indicating that the set $\|\tilde{\bm{d}}\|\leq d_m/c_1$ is attractive and forward invariant. Therefore, $\tilde{\bm{d}}(t)$ remains within the set $\|\tilde{\bm{d}}\|\leq \tilde{d}_m$ for all $t\geq0$. As per Lemma \ref{L_PPTS}, it follows from the first inequality of \eqref{dV_d1} that the PTDO is globally PPTS, in the sense that $\tilde{\bm{d}}$ converges to a small residual set within the prescribed time $T_{ce}$.  $\hfill \blacksquare$

In the following, the PTDO is used to design the reduced-attitude tracking controller. Since a safety margin of size $\varepsilon$ is incorporated into the design of boresight guidance law, if the condition $\text{d}_{\mathbb{S}^{2}}(\bm{\sigma}(t),\bm{b})<\varepsilon$ (equivalent to $\text{d}_{\mathbb{S}^{2}}(\bm{x}(t),\bm{x}_r(t))<\varepsilon$) holds for all $t\geq0$, then the inertial-frame boresight vector $\bm{x}(t)$ remains outside all forbidden pointing zones. To ensure that $\text{d}_{\mathbb{S}^{2}}(\bm{\sigma}(t),\bm{b})<\varepsilon$ for all $t\geq0$, the following constraint is imposed on $\sigma_e$:
\begin{equation}
\label{tube}
\sigma_e(t) < \rho:=1 - \cos\varepsilon, ~\forall t\geq0.
\end{equation}
This implies that, at each moment, the boresight vector $\bm{x}(t)$ is allowed to stay within a conical zone centered around the axis $\bm{x}_r(t)$, with a half-angle of $\varepsilon$. Unifying all such conical zones along $t$ forms a fixed-size ``safe tube'' around $\bm{x}_r(t)$. To deal with this constraint, we define a transformed error
\begin{equation}
\label{xi}
\xi(t) := \dfrac{\sigma_e(t)}{\rho}
\end{equation}
whose governing equation is derived from \eqref{dx_e} as 
\begin{equation}
\label{dxi}
\dot{\xi}(t) = -\dfrac{1}{\rho}\bm{b}^\top[\bm{\sigma}]_\times \bm{\omega}_e.
\end{equation}
It is evident from \eqref{xi} that $0\leq\xi(t)<1$ is equivalent to \eqref{tube}, and $\xi(t)=0$ only when $\sigma_{e}(t)=0$. Therefore, the boresight tracking problem boils down to achieving practical prescribed-time convergence of $\xi(t)$, while ensuring $0\leq\xi(t)<1$ holds for all $t\geq0$. 

Let us define $\bm{z}:=\bm{\omega}_e-\bm{\omega}_c$,
where $\bm{\omega}_c\in\mathbb{R}^3$ is a virtual control law to be designed. Consider a barrier function $V_\xi=\ln \frac{1}{1-\xi}$. Taking its time derivative along \eqref{dxi} gets
\begin{align}
\label{dV_xi}
\dot{V}_\xi=-\dfrac{1}{\rho(1-\xi)}\bm{b}^\top[\bm{\sigma}]_\times (\bm{z}+\bm{\omega}_c).
\end{align}
Design the virtual control law as
\begin{align}
\label{omega_c}
\bm{\omega}_c=-c_2\mu_c(t)[\bm{\sigma}]_\times\bm{b}
\end{align}
where $c_2>0$ is a constant. It is noted that $\mu_c(t)$ and $\bm{\sigma}$ are continuously differentiable, so does $\bm{\omega}_c$. Then, substituting  \eqref{omega_c} into \eqref{dV_xi}, whilst recalling \eqref{err}, we get
\begin{align}
\label{dV_xi_1}
\dot{V}_\xi&= -\dfrac{c_2\mu_c(t)}{\rho(1-\xi)}(1 - (\bm{b}^\top\bm{\sigma})^2)-\dfrac{\bm{b}^\top[\bm{\sigma}]_\times\bm{z}}{\rho(1-\xi)} \nonumber\\
&\leq -c_2(2+\rho)\mu_c(t) V_\xi -\dfrac{\bm{b}^\top[\bm{\sigma}]_\times\bm{z}}{\rho(1-\xi)}
\end{align}
where the fact that $1 + \bm{b}^\top\bm{\sigma}\geq 2+\rho$ in the set $0\leq\xi(t)<1$ has been used.

In view of \eqref{domega}, the time derivative of $\bm{z}$ is derived as
\begin{align}
\label{dz}
\bm{J}\dot{\bm{z}}= - \bm{J}\dot{\bm{\omega}}_c + \bm{H} + \bm{u} + \bm{d}.
\end{align}
Now design the reduced-attitude control law as
\begin{equation}
\label{u}
\bm{u} = - c_3\mu_c(t)\bm{z} + \bm{J}\dot{\bm{\omega}}_c - \bm{H} -\hat{\bm{d}} - \dfrac{[\bm{\sigma}]_\times \bm{b}}{\rho(1-\xi)}
\end{equation}
where $c_3>0$ is a constant. Substituting \eqref{u} into \eqref{dz} gets
\begin{align}
\label{dz1}
\bm{J}\dot{\bm{z}}= - c_3\mu_c(t)\bm{z} + \tilde{\bm{d}} - \dfrac{[\bm{\sigma}]_\times \bm{b}}{\rho(1-\xi)}.
\end{align}
To facilitate the subsequent analysis, we define the closed-loop error as $\bm{Z}:= [\xi,\bm{z}^\top,\tilde{\bm{d}}^\top]^\top\in\mathbb{R}^{7}$.

\begin{theorem} \label{Th2}
	Consider the reduced-attitude dynamics given by \eqref{domega} and \eqref{dx} under Assumption \ref{A1}. The tracking controller \eqref{u}, along with the PTDO defined by \eqref{dp} and \eqref{hat_d}, ensures the following:
	
	\begin{enumerate}	
		
		\item All closed-loop signals are bounded.
		
		\item The reduced-attitude tracking error $\sigma_e$ complies with the constraint defined by \eqref{tube} at all times.
		
		\item The closed-loop system is PPTS, in the sense that the error vector $\bm{Z}$ converges to a small residual set within the prescribed time $T_{c}^\ast$. 
	\end{enumerate}
\end{theorem}

\textit{Proof}. Consider the Lyapunov function candidate
\begin{equation}
V = V_d + V_\xi +  \dfrac{1}{2}\bm{z}^\top\bm{J}\bm{z}
\end{equation}
Taking the time derivative of $V$ along \eqref{dV_d1}, \eqref{dV_xi_1}, and \eqref{dz1} gets
\begin{align}
\label{dV}
\dot{V} 
\leq & - c_1 \mu_c(t) V_d -c_2(2+\rho)\mu_c(t) V_\xi - \dfrac{c_3\mu_c(t)}{2}\|\bm{z}\|^2   \nonumber \\
& + \dfrac{1}{2c_3}\sup_{t\geq0}\|\tilde{\bm{d}}(t)\|^2 + \dfrac{1}{2c_1} d_m^2 \nonumber \\
\leq &- \alpha\mu_c(t)V + \beta
\end{align}
where $\alpha:=\min\{c_1,c_2(2+\rho),\frac{c_3}{\lambda_{M}(\bm{J})}\}>0$ with $\lambda_{M}(\bm{J})$ being the maximum eigenvalue of $\bm{J}$, and $\beta:= \frac{1}{2c_1} d_m^2 + \frac{1}{2c_3}\tilde{d}_m^2 >0$. From \eqref{dV}, it follows that $V$ is bounded according to Lemma \ref{L_PPTS}. Then, one can conclude that $V_\xi$ and all closed-loop signals are bounded. The boundedness of $V_\xi$ implies that $0\leq\xi(t)<1$ for all $t\geq 0$, and consequently, the tracking error $\sigma_e$ respects the constraint \eqref{tube}. According to Lemma \ref{L_PPTS}, we conclude from \eqref{dV} that the error vector $\bm{Z}$ converges to a residual set within the prescribed time $T_{c}^\ast$. In particular, the set size can be made arbitrarily small by increasing $T_{c}^\ast$ closer to $T_c$.  $\hfill \blacksquare$

\begin{remark} \label{R2}
	The proposed PTDO-based tracking controller \eqref{u} is capable of tackling inertia uncertainties, by expressing the inertia matrix as $\bm{J}=\bm{J}_0+\Delta \bm{J}$, where $\bm{J}_0$ is the nominal inertia matrix, and $\Delta \bm{J}$ denotes the uncertain components. The system dynamics \eqref{Jd_omega} can then be rewritten as: 
	\begin{equation}
	\nonumber
	\bm{J}_0\dot{\bm{\omega}}_e =-\bm{C}_0\bm{\omega}_e - \bm{G}_0 + \bm{u} + \bm{d}_l
	\end{equation}
	where $\bm{C}_0$ and $\bm{G}_0$ represent the nominal parts of $\bm{C}$ and $\bm{G}$, respectively; while $\bm{d}_l$ is a lumped disturbance that captures  the external disturbance $\bm{d}$ and all unknown terms associated with $\Delta \bm{J}$. The proposed PTDO can effectively reconstruct $\bm{d}_l$ within a prescribed time,  as demonstrated by the experimental results in Sec. \ref{secV}.
\end{remark}

\begin{remark}
	Under the proposed IBGC framework, if the tube radius $\rho$ is less than the safety margin $\epsilon$ and the controller settling time $T_{c}^\ast$ does not exceed the guidance time $T_{g}^\ast$, then the boresight vector $\bm{x}$ will converge to a small neighborhood of the desired orientation $\bm{x}^\ast$ within $T_{g}^\ast$, while avoiding all forbidden zones $\mathcal{P}_i$, $i\in\mathbb{I}$. Therefore, the proposed IBGC scheme achieves pointing-constrained boresight reorientation within a preassigned time tailored to specific task demands, in the presence of external disturbances.
\end{remark}

\section{Illustrative example} \label{secIV}

In this section, a numerical example is provided to show the efficacy of the proposed IBGC scheme. The spacecraft inertia vector is $\bm{J}=[20,1.2,0.9; 1.2,17,1.4;0.9,1.4,15]\text{kg}\cdot\text{m}^2$. The spacecraft is required to reorient the boresight vector $\bm{x}$ (its expression in the body-fixed frame is $\bm{b}=[0,0,1]^\top$) of the onboard light-sensitive telescope from an initial pointing to the desired orientation $\bm{x}^\ast=[-0.939,-0.305,0.1589]^\top$. The disturbance $\bm{d}$ is modeled as
\begin{equation}
\nonumber
\bm{d}=10^{-3}\times\left[
\begin{matrix}
3\cos(0.2t)+4\sin(0.06t)-1     \\
-1.5\sin(0.04t)+3\cos(0.1t)+1.5  \\
3\sin(0.2t)-8\sin(0.08t)+1.5
\end{matrix}\right]\rm Nm.
\end{equation}
During this maneuver, the boresight vector $\bm{x}$ needs to avoid certain forbidden zones, with parameters listed in Table \ref{tab1}. The task completion time is $150\,\rm s$. 

\begin{figure}[!htbp]
	\centering
	\subfigure[{Viewpoint 1}]{
		\includegraphics[width=3.85cm]{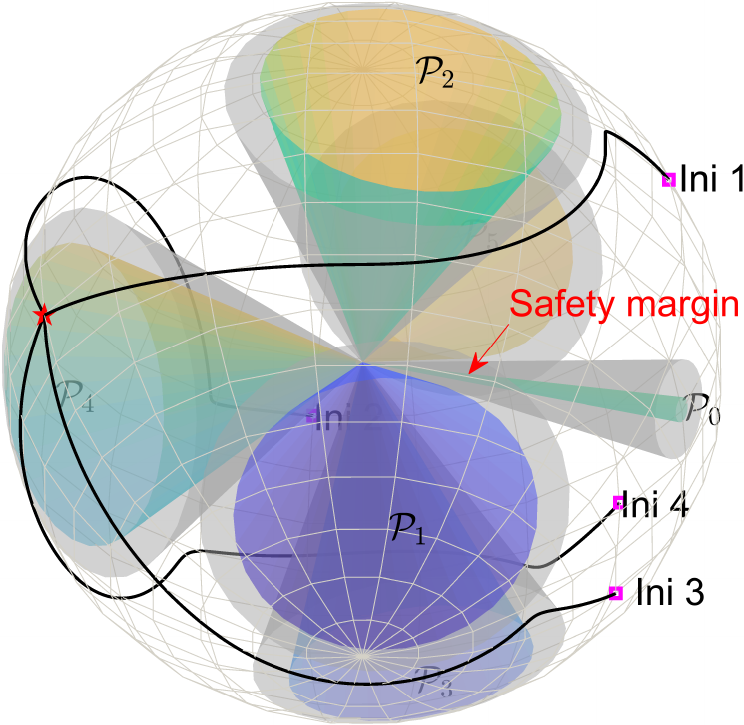}
		\label{guidance_3D_view1}}
	\subfigure[{Viewpoint 2}]{
		\includegraphics[width=3.8cm]{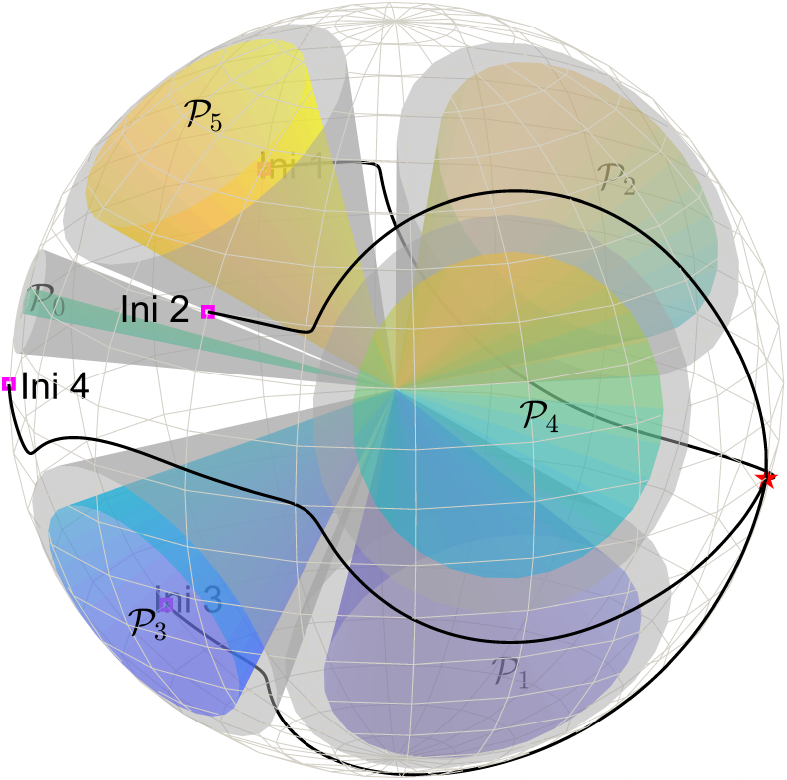}
		\label{guidance_3D_view2}}
	\caption{Boresight trajectories on $\mathbb{S}^2$ in the inertial frame, where the purple squares denote different initial orientations and the red star is the desired orientation $\bm{x}^\ast$.}
	\label{Guidance_trajectory}
\end{figure}

\begin{figure}[!htbp]
	\centering
	\subfigure[{PPT-BG}]{
		\includegraphics[width=8.8cm]{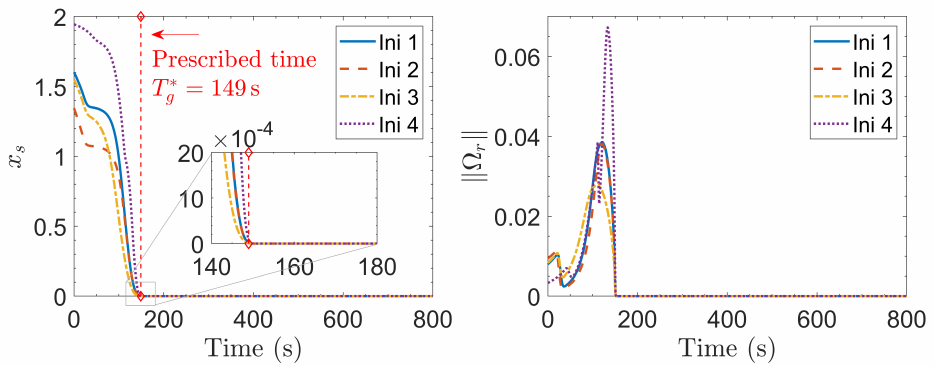}
		\label{error_PPT_BG}}
	\subfigure[{APF-BG}]{
		\includegraphics[width=8.8cm]{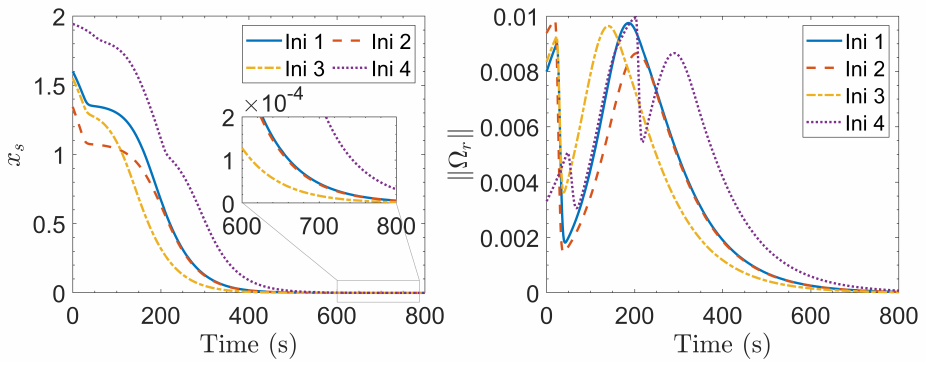}
		\label{error_APF_BG}}
	\caption{Comparison of the responses of PPT-BG and APF-BG under four distinct initial orientations labeled as `Ini 1' to `Ini 4'.}
	\label{Guidance_error}
\end{figure}

We begin by verifying the effectiveness of the prescribed-time guidance law $\bm{\Omega}_r=\bm{h}_g(\bm{x}_r,t)$ in \eqref{h_g}, referred to as `PPT-BG'. For comparison, the baseline guidance law $\bm{\Omega}_r=\bm{h}(\bm{x}_r)$ in \eqref{h}, denoted as `APF-BG', is also simulated. The design parameters are tabulated in Table \ref{tab2}. The guidance trajectories starting from a set of initial orientations in the inertial frame are depicted in Fig. \ref{Guidance_trajectory}. As observed, the PPT-BG successfully guides the boresight vector $\bm{x}_r$ from various initial orientations to the goal, while avoiding all forbidden zones. Note that the APF-BG follows the same path as the PPT-BG, and thus its trajectory is not shown here. To quantify the alignment, let us define the boresight error as $x_s:= 1 - \bm{x}_r^\top \bm{x}^\ast$. The comparison results of $x_s$ and $\|\bm{\Omega}_r\|$ are depicted in Fig. \ref{Guidance_error}, from which it is evident that the PPT-BG achieves convergence of $\bm{x}_s$ to a very small neighborhood of the origin within the prescribed time $T_g^\ast=149\,\rm s$, regardless of initial orientations (see the left subfigure of Fig. \ref{Guidance_error}(a)). In contrast, the APF-BG only achieves asymptotic error convergence, that is, $\bm{x}_r$ converges to zero as time goes to infinity, as shown in the left subfigure of Fig. \ref{Guidance_error}(b); moreover, the convergence time increases with the initial error. This result highlights the crucial role of the PPTA function $\mu_g(t)$ in enabling practical prescribed-time convergence of $x_s$. Note that a faster convergence rate is accompanied by a larger velocity norm, as evident in the left subfigures of Fig. \ref{Guidance_error}.

\begin{table}
	\caption{Geometrical details of forbidden zones. \label{tab1}}
	\begin{threeparttable}
		\begin{tabular*}{250pt}{@{\extracolsep\fill}cll@{\extracolsep\fill}}%
			\toprule
			Forbidden zone   &       Central axis        &       Angle       \\
			\midrule
			$\mathcal{P}_0$                  &   $\bm{f}_0=-\bm{x}^\ast$     &    $\theta_0=2^{\circ}$    \\
			
			$\mathcal{P}_1$                   &   $\bm{f}_1=[0, -0.453,-0.8915]^{\top}$          &    $\theta_1=25^{\circ}$    \\ 
			
			$\mathcal{P}_2$                   &   $\bm{f}_2=[0, -0.951,0.3092]^{\top}$    &    $\theta_2=25^{\circ}$    \\
			
			$\mathcal{P}_3$                   &   $\bm{f}_3=[0.275, 0.847,-0.4549]^{\top}$    &    $\theta_3=20^{\circ}$    \\
			
			$\mathcal{P}_4$                   &   $\bm{f}_4=[-0.769, 0.599,0.2232]^{\top}$    &    $\theta_3=25^{\circ}$    \\
			
			$\mathcal{P}_5$                   &   $\bm{f}_5=[0.345, 0.475,0.8095]^{\top}$    &    $\theta_3=20^{\circ}$    \\
			\bottomrule
		\end{tabular*}
	\end{threeparttable}
\end{table}

\begin{table}
	\caption{Guidance and control parameters.\label{tab2}}
	\centering
	\begin{threeparttable}
		\begin{tabular*}{250pt}{@{\extracolsep\fill}lcc@{\extracolsep\fill}}%
			\toprule
			Method      &     Parameter     \\
			\midrule
			PPTA function $\mu_g(t)$ &   $T_g=150$, $T_{g}^\ast=149$                       \\
			PPTA function $\mu_c(t)$ & $T_{c}=15$, $T_{c}^\ast=14$         \\
			PPT-BG in \eqref{h_g}        &     $\epsilon=\frac{\pi}{30}$, $\epsilon^\ast=\frac{\pi}{12}$, $k_a=0.01$, $k_r=0.1$               \\
			PPT-TC in \eqref{u} &   $c_1=c_2=c_3=0.2$, $\rho=\epsilon$, $\bm{p}(0)=\bm{0}_3$       \\
			APF in \cite{hu2019reduced}        &     $k_p=5$, $k_d=2$             \\ 
			PD in \cite{chaturvedi2011rigid}         &     $k_p=0.05$, $k_d=2$             \\  
			\bottomrule
		\end{tabular*}
	\end{threeparttable}
\end{table}

\begin{figure}[hbt!]
	\centering
	\includegraphics[width=8.8cm]{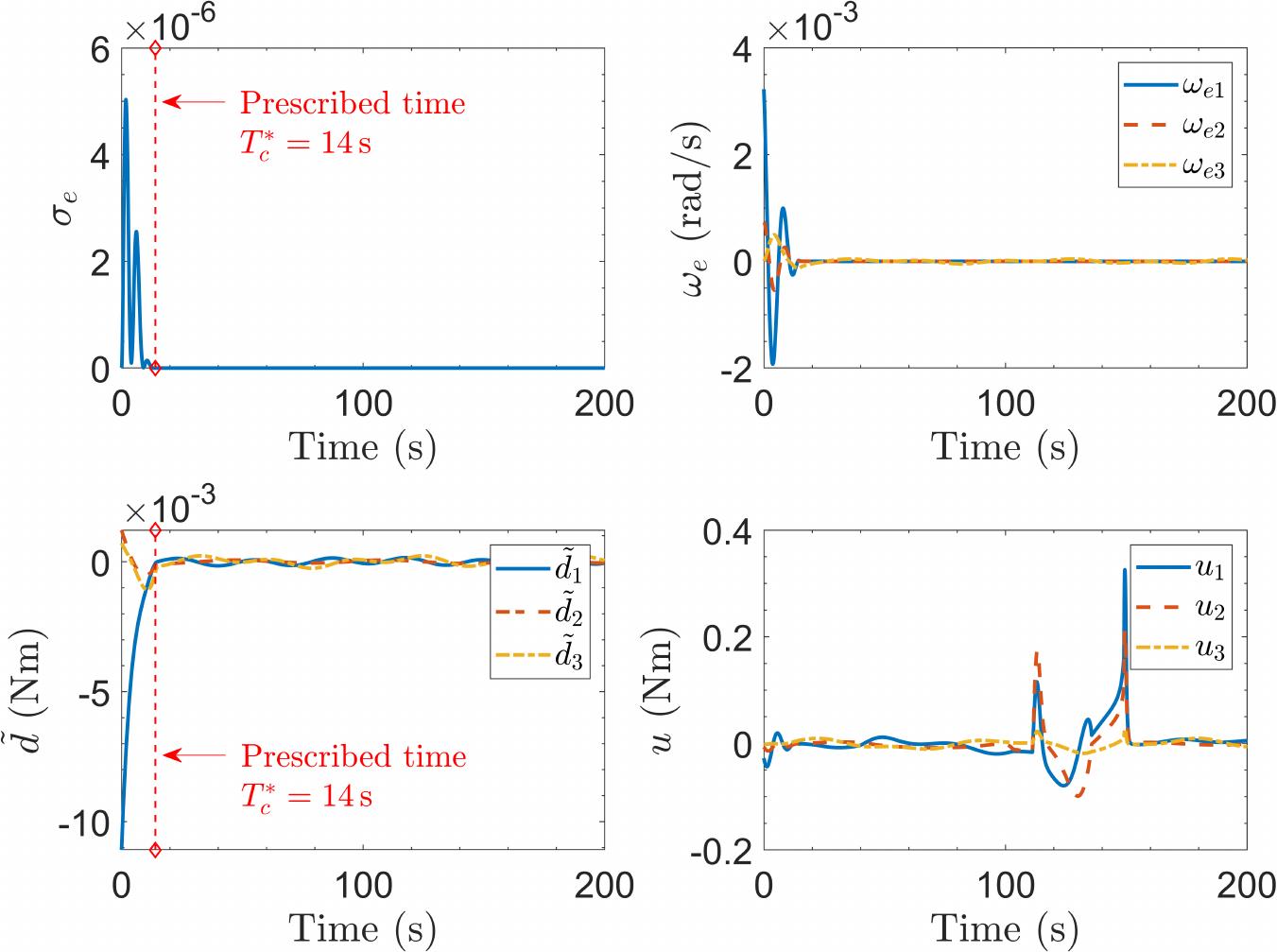}
	\caption{Control responses of the proposed PPT-TC.}
	\label{control_response}
\end{figure}

\begin{figure}[hbt!]
	\centering
	\includegraphics[width=6.5cm]{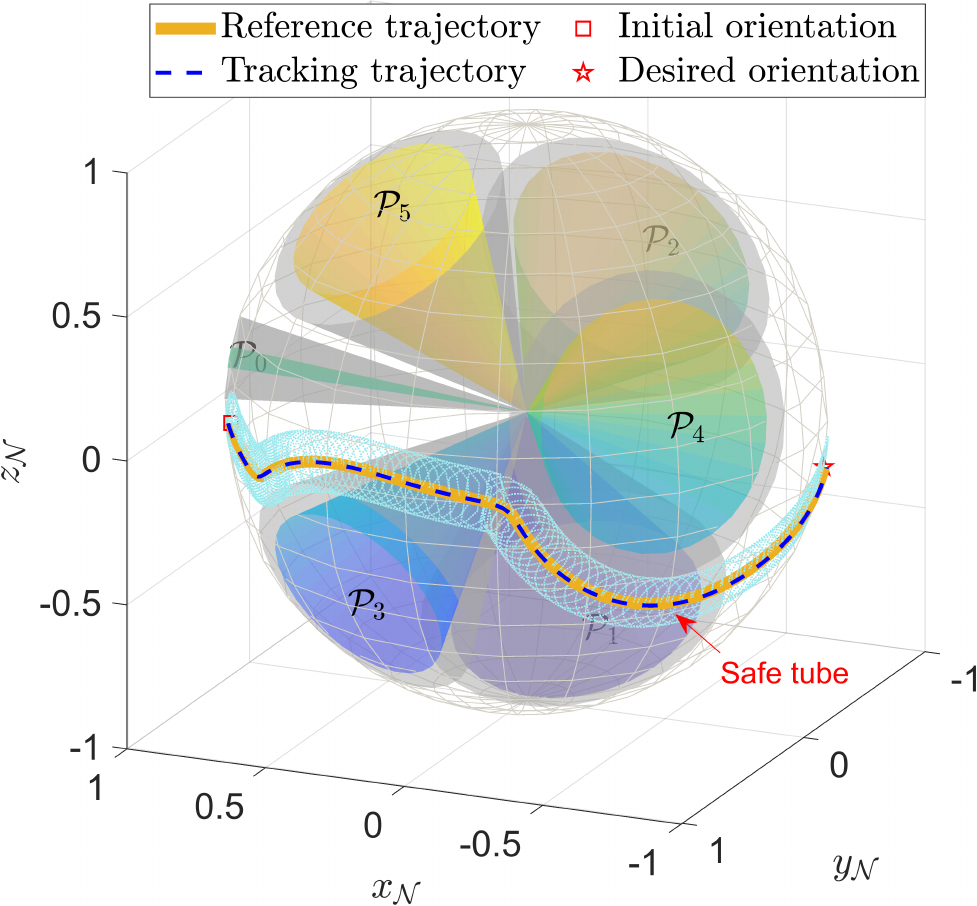}
	\caption{Boresight tracking trajectory on $\mathbb{S}^2$.}
	\label{3D_tracking}
\end{figure}

We next verify the proposed reduced-attitude tracking controller \eqref{u}, referred to as `PPT-TC'. The controller parameters are detailed in Table \ref{tab2}. As a case study, the guided trajectory starting from the 4th initial orientation is selected as the reference trajectory. The closed-loop responses are shown in Fig. \ref{control_response}. As can be seen, the proposed PPT-TC ensures that the output-tracking errors $\sigma_e$ and $\bm{\omega}_e$, as well as the disturbance estimation error $\tilde{\bm{d}}$, converge to small residual sets within the prescribed time $T_{c}^\ast = 14\,\rm s$; moreover, it maintains $\sigma_e$ within the preset ``tube'' with a radius of $\rho = \epsilon$. Consequently, the boresight vector $\bm{x}$ reaches a small neighborhood around the desired orientation $\bm{x}^\ast$ within the prescribed time $T_{g}^\ast = 149\,\rm s$, while avoiding all forbidden zones, as shown in Fig. \ref{3D_tracking}. This ensures successful completion of the boresight reorientation task within a required time of $150\,\rm s$. 

%

\section{Experimental Results} \label{secV}

\begin{figure}[hbt!]
	\centering
	\includegraphics[width=8cm]{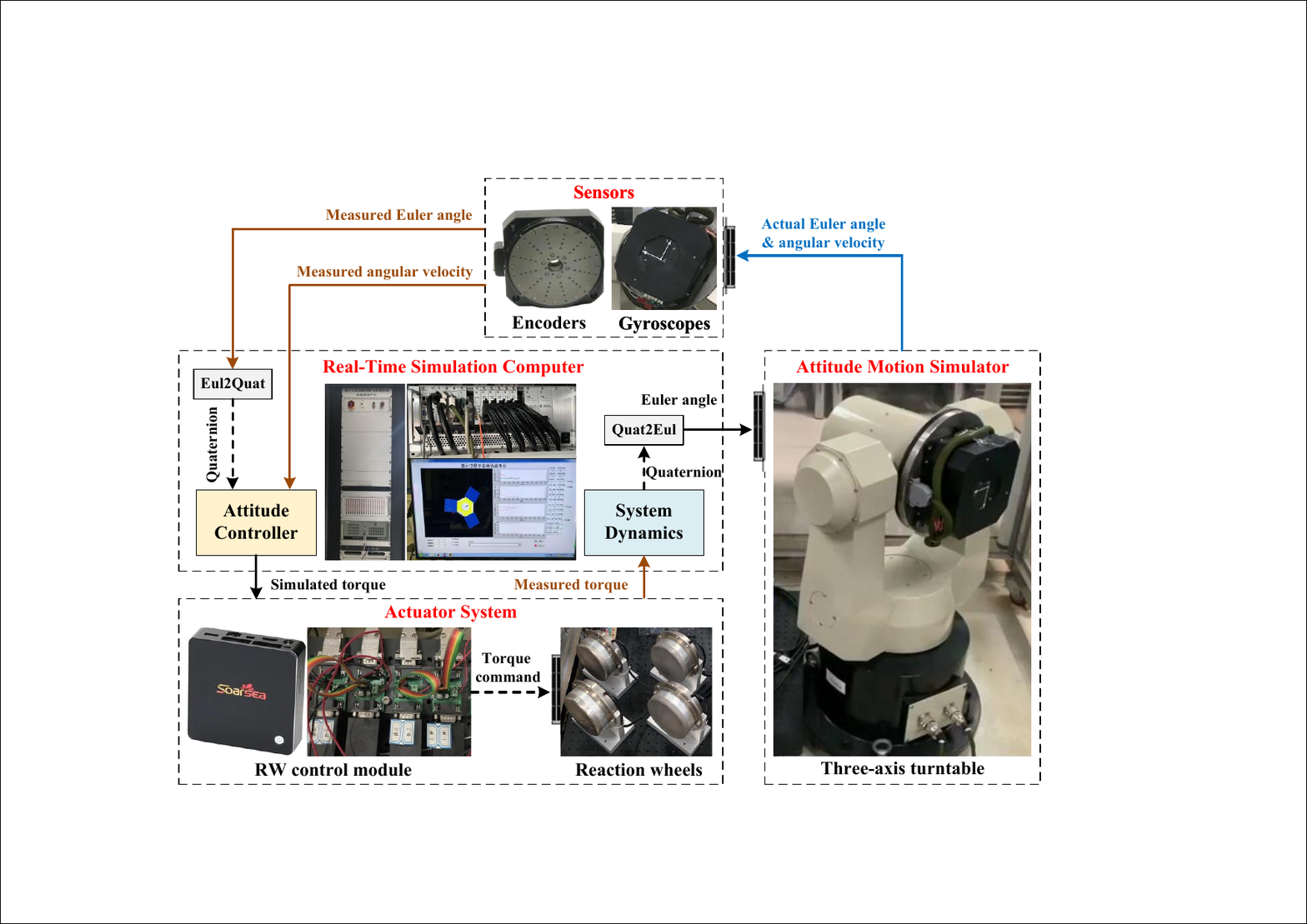}
	\caption{Schematic of the HIL testbed.}
	\label{HIL_setup}
\end{figure}

In this section, the hardware-in-the-loop (HIL) experiments are performed to demonstrate the practical effectiveness of the proposed IBGC scheme. The experimental setup is provided in Fig. \ref{HIL_setup}. The testbed consists of the following main modules: 1) a high-performance simulation computer (HPRTSC), which runs the attitude dynamics on the VxWorks system; 2) a three-axis turntable for simulating the spacecraft attitude motion from the dynamics in HPRTSC; 3) a set of sensors including three grating encoders and four fiber-optic gyroscopes, which are mounted on the turntable and utilized for measuring Euler angles and angular velocities; 4) an actuator system consisting of four reaction wheels (RWs) and a control module that receives the torque signals from the controller in the HPRTSC and allocates corresponding torque commands to RWs. Each RW provides a maximum torque of $0.1\,\text{Nm}$ and a maximum speed of $5000\,\text{rpm}$. 

To avoid the singularity that occurs when the pitch angle equals $90^\circ$ (caused by the structure limitations of turntable), the initial value of the boresight vector is selected as $\bm{x}(0)=[0.809,0.587,0.0308]^\top$ (i.e., Ini 2 in Fig. \ref{Guidance_trajectory}). In addition, the nominal inertia matrix and the disturbance $\bm{d}$ are given in Sec. \ref{secIV}. The inertia uncertainties caused by phenomena, such as payload motion and appendage deployment, are set to $\Delta \bm{J} = \text{diag}[-3\tanh(0.1 t) -1, 2\sin(0.05t)+3,\cos(0.1t)+3]\,\text{kg}\cdot\text{m}^2$, which injects time-varying inertia uncertainties into the closed-loop systems.

Two classical reduced-attitude control methods are compared with the proposed IBGC method. The first is the APF-based controller introduced in \cite{hu2019reduced} (referred to as `APF'), which is defined as $\bm{u}=-k_p\bm{R}^\top[\bm{x}]_\times \nabla U(\bm{x}) - k_d \bm{\omega}$. The second is the PD control introduced in \cite{chaturvedi2011rigid} (referred to as `PD'), expressed as $\bm{u}=k_p\bm{R}^\top[\bm{x}]_\times \bm{x}_d - k_d \bm{\omega}$,
where $k_p,\,k_d>0$ are constant gains with values given in Table \ref{tab2}. 

\begin{figure}[hbt!]
	\centering
	\includegraphics[width=8.8cm]{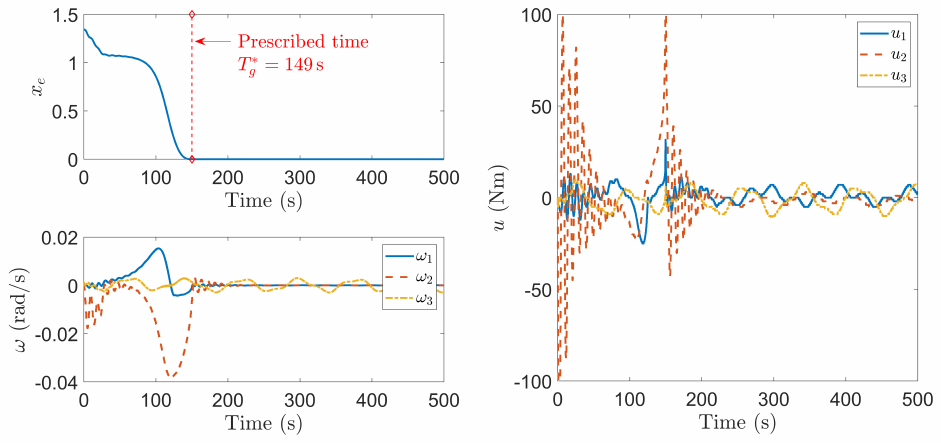}
	\caption{Experimental responses of the proposed IBGC scheme.}
	\label{Dynamic_responses_expe}
\end{figure}

\begin{figure}[hbt!]
	\centering
	\includegraphics[width=8.8cm]{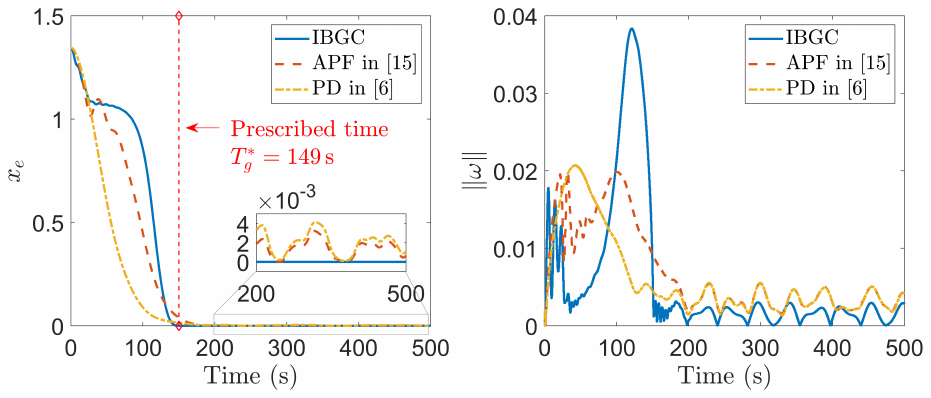}
	\caption{Comparison of experimental results.}
	\label{Comparision_results_expe}
\end{figure}

\begin{figure}[hbt!]
	\centering
	\includegraphics[width=6.5cm]{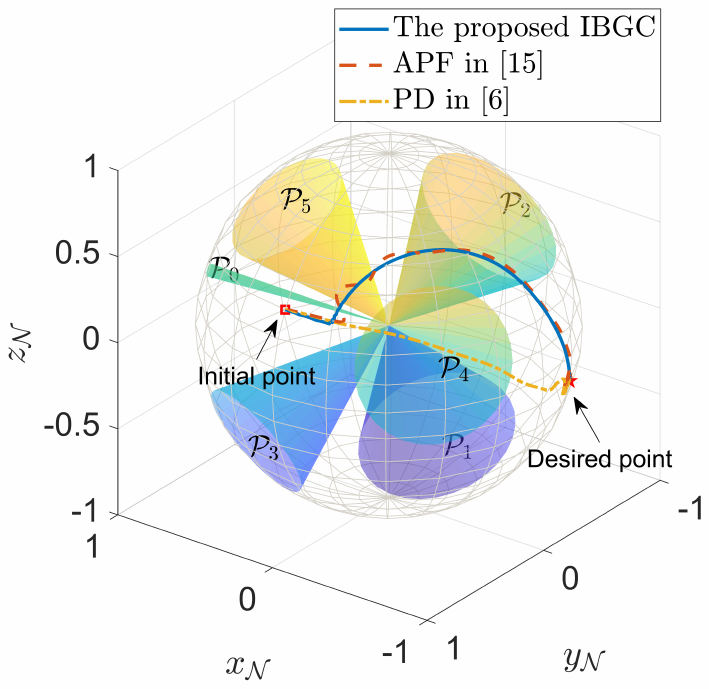}
	\caption{Comparison of 3D boresight trajectories.}
	\label{3D_trajectories_expe}
\end{figure}

The experimental results of the proposed IBGC scheme are shown in Fig. \ref{Dynamic_responses_expe}. It is clear that the boresight error, defined as $x_e:= 1 - \bm{x}^\top \bm{x}^\ast$, converges to a very small neighborhood of the origin within the prescribed time $T_g^\ast=149\,\rm s$. This implies that the boresight vector $\bm{x}$ achieves the desired orientation within the required time of $150\,\rm s$. Furthermore, the torque outputs of the RWs remain within the magnitude limits. The comparison results are plotted in Fig. \ref{Comparision_results_expe}. As can be seen, both the APF and PD methods fail to achieve boresight reorientation within the required time of $150\,\rm s$. While it is possible to improve their convergence time by increasing the control gains, this requires empirical parameter tuning. In contrast, the proposed IBGC scheme allows for explicit pre-specification of convergence time without dependence on gain selection, making it more suitable for practical applications. Furthermore, from the left subfigure of Fig. \ref{Comparision_results_expe}, one can observe that our scheme exhibits highest accuracy among these three methods. As dictated by Remark \ref{R2}, the proposed IBGC scheme has strong robustness against both external disturbances and inertia uncertainties. This robustness stems from the incorporation of a PTDO (defined by \eqref{dp} and \eqref{hat_d}), which enables the proposed IBGC scheme to achieve high-precision boresight control in the perturbed and uncertain scenarios. While both the simulated APF and PD controllers are inertia-independent, their performance remains limited in the perturbed scenarios, due to the lack of disturbance compensation mechanisms. The 3D boresight trajectories for these three methods are presented in Fig. \ref{3D_trajectories_expe}, from which it is clear that both IBGC and APF successfully avoid the forbidden pointing zones, whereas the PD controller fails. In summary, the experimental results demonstrate the practical effectiveness of the proposed IBGC scheme in terms of prescribed time and pointing avoidance. A video demonstrating the experimental results is available via the link \url{https://youtu.be/mhHY0x6dFAE}.

\section{Conclusions and Future Works} \label{secVI}

This article develops a PPTA function and integrates it into the IBGC scheme, providing a systematic and flexible control framework to address the boresight orientation problem on $\mathbb{S}^2$ under temporal and pointing constraints. The salient features of the proposed method, as demonstrated through simulation and experimental results, are two-fold:  1) it achieves boresight reorientation within a finite time that can be arbitrarily preassigned according to the required task completion time, while ensuring that the boresight vector avoids all forbidden pointing zones; 2) it enables precise reconstruction of unknown disturbances within a prescribed time. For future work, we envisage extending our method to multi-spacecraft attitude coordination with explicit consideration of tempo-spatial constraints.

\section*{Appendix A} 
\label{Appendix_A}

The proof is analyzed over two time intervals: $[0,T^\ast)$ and $[T^\ast,\infty)$. Consider the first time interval, in which $\mu(t)=\mu_p(t)$. Multiplying both sides of \eqref{dV_PTS} by the integrating factor $(T-t)^{-\alpha T}$ yields
\begin{align}
\nonumber
(T-t)^{-\alpha T}\dfrac{dV}{dt} + \alpha(T-t)^{-\alpha T}\mu(t)V\leq \beta(T-t)^{-\alpha T}.
\end{align}
whose left-hand side can be expressed as
\begin{align}
\label{dV_PTS_2}
\dfrac{d}{dt}[(T-t)^{-\alpha T}V]\leq \beta (T-t)^{-\alpha T}.
\end{align}
Integrating both sides of \eqref{dV_PTS_2} from $0$ to $T_e$ yields
\begin{align}
\nonumber
(T-T^\ast)^{-\alpha T}&V(\bm{y}(T^\ast))\leq T^{-\alpha T}V_0  \nonumber\\
&+\dfrac{\beta}{\alpha T-1}\left[(T-T^\ast)^{1-\alpha T} -T^{1-\alpha T} \right].
\end{align}
Therefore, the solution of the system \eqref{dy} converges to the set $V(\bm{y})\leq v_1$ at the prescribed time $t=T^\ast$. 

We next consider the time interval $[T^\ast,\infty)$, where $\mu(t)\geq \mu(T^\ast)>0$. From \eqref{dV_PTS}, it follows that 
\begin{equation}
\label{dV_T_e}
\dot{V}\leq - \alpha_s V+ \beta
\end{equation}
with $\alpha_s:=\alpha\mu(T^\ast)$. Then, one can easily deduce that
\begin{equation}
\label{V_T_e}
V(\bm{y}(t))\leq\left(V(\bm{y}(T^\ast))-\dfrac{b}{\alpha_s}\right)e^{-\alpha_s(t-T^\ast)} + \dfrac{\beta}{\alpha_s}.
\end{equation}

To proceed, two cases are discussed. 

Case 1: $V(\bm{y}(T^\ast))>v_2$. It follows from \eqref{V_T_e} that $V(\bm{y}(t))$ decreases exponentially to the set $V(\bm{y})\leq v_2$. This, together with the fact that $V(\bm{y}(T^\ast))\leq v_1$, ensures that $\bm{y}(t)$ remains within the set $V(\bm{y})\leq v_1$ for all $t\geq T^\ast$.

Case 2: $V(\bm{y}(T^\ast))\leq v_2$. Inspecting \eqref{dV_T_e} finds that $\dot{V}(\bm{y})=0$ on the level set $V(\bm{y})=v_2$. This implies that the set $V(\bm{y})\leq v_2$ is forward invariant. Thus, from the fact that $V(\bm{y}(T^\ast))\leq v_2$, it follows that $V(\bm{y}(t))\leq v_2$ for all $t\geq T^\ast$. 

With the above two cases in mind, one can easily conclude that $\bm{y}(t)$ remains within the set $\Omega$ given by \eqref{Omega} for all $t\geq T^\ast$. Furthermore, given that $T-T^\ast\ll \alpha T-1$, it is evident from \eqref{V_e} that the size of the set $\Omega$ can be artificially reduced by increasing $T^\ast$ closer to $T$. According to Definition \ref{D_PTS}, one can conclude that the origin of \eqref{dy} is PPTS. If $\mathbb{U}=\mathbb{R}^n$, it is easy to show the globally practical prescribed-time stability using radially unbounded Lyapunov functions. $\hfill \blacksquare$

\section*{Appendix B}  \label{Appendix_B}

Taking the time derivative of $U(\bm{x}_r)$ along \eqref{dx_h} yields
\begin{align}
\dot{U}(\bm{x}_r) &= -\nabla U(\bm{x}_r)^\top[\bm{x}_r]_\times[\bm{x}_r]_\times^\top \nabla U(\bm{x}_r) \nonumber \\
&= -\|[\bm{x}_r]_\times^\top \nabla U(\bm{x}_r) \|^2 \leq 0 \nonumber
\end{align}
which implies that $U(\bm{x}_r)$ is bounded. Thus, we have $\bm{x}_r(t)\in\mathcal{F}_\epsilon$ for all $t\geq0$ and $\bm{x}_r(0)\in\mathcal{F}_\epsilon$, that is, the practical free space $\mathcal{F}_\epsilon$ is forward invariant. Applying LaSalle's invariance principle, we further conclude that the solution of \eqref{dx_h} asymptotically converges to the set of critical points, given by
\begin{equation}
\nonumber
{\rm Crit}~U:=\left\lbrace \bm{x}_r\in\mathbb{S}^2 \, \big| \, [\bm{x}_r]_\times^\top \nabla U(\bm{x}_r)=\bm{0}\right\rbrace .
\end{equation}
It is easy to verify that $[\bm{x}_r]_\times^\top \nabla U(\bm{x}_r)=\bm{0}$ holds if and only if $\bm{x}_r$ and $\nabla U(\bm{x}_r)$ are collinear, or $\nabla U(\bm{x}_r)=\bm{0}$. Given that $\bm{x}^\ast\in\mathbb{S}^2\setminus\bigcup_{i=0}^{m}\mathcal{I}_i$ and $\nabla\phi_i(\bm{x}_r^\top\bm{f}_i)\geq0$, it follows from \eqref{nabla_U} that $\nabla U(\bm{x}_r)\neq\bm{0}$, thus precluding the latter condition. Next, we consider two cases to analyze the former one.

\textit{Case 1:} $\bm{x}_r\in\mathbb{S}^2\setminus\bigcup_{i=0}^{m}\mathcal{I}_i$, indicating that $\bm{x}_r$ remains outside the influence regions of all forbidden pointing constraints. In this case, we have $\nabla\phi_i(\bm{x}_r^\top\bm{f}_i)=0$ for all $i\in\mathbb{I}$. Then, the gradient $\nabla U(\bm{x}_r)$ given by \eqref{nabla_U} reduces to $\nabla U(\bm{x}_r)=-k_a\bm{x}^{\ast}$. Solving $[\bm{x}_r]_\times^\top \nabla U(\bm{x}_r)=k_a[\bm{x}_r]_\times \bm{x}^\ast=\bm{0}$ yields $\bm{x}_r=\pm\bm{x}^\ast$. However, since $-\bm{x}^\ast\in\mathcal{P}_0$ by construction, the only feasible solution is the desired equilibrium $\bm{x}_r=\bm{x}^\ast$.

\begin{figure}[!htbp]
	\centering
	\subfigure[{Forbidden zone $\mathcal{P}_0$}]{
		\includegraphics[width=3.2cm]{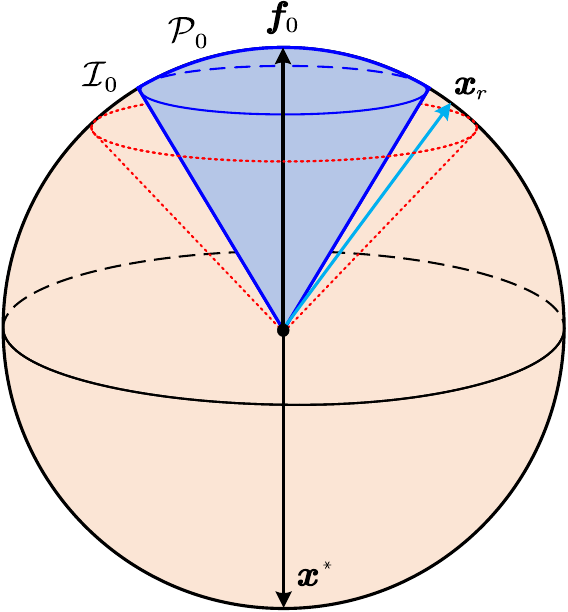}
		\label{Critical_points_1}}
	\subfigure[{Forbidden zone $\mathcal{P}_i$, $i\neq0$}]{
		\includegraphics[width=3.45cm]{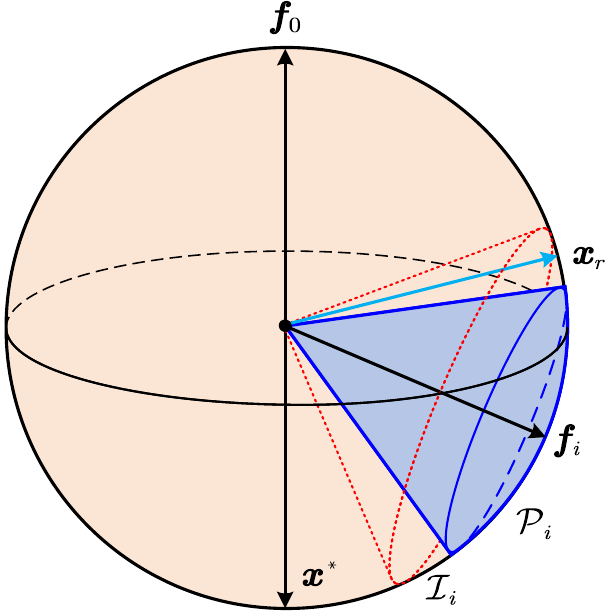}
		\label{Critical_points_2}}
	\caption{Illustration of potential critical points for $i$-th forbidden zone.}
	\label{Critical_points}
\end{figure}

\textit{Case 2:} $\bm{x}_r\in\mathcal{I}_i$, $i\in\mathbb{I}$. As the influence regions of pointing constraints are pairwise disjoint, it is sufficient to examine the critical points for each pointing constraint $\mathcal{P}_i$ separately. Then, $[\bm{x}_r]_\times^\top \nabla U(\bm{x}_r)=\bm{0}$ is equivalent to \eqref{Crit}. As $\bm{x}^\ast\in\mathbb{S}^2\setminus\bigcup_{i=0}^{m}\mathcal{I}_i$ and $-\bm{x}^\ast\in\mathcal{P}_0$, it is evident that $[\bm{x}_r]_\times\bm{x}^{\ast}\neq \bm{0}$ for $\bm{x}_r\in\mathcal{I}_i$, $i\in\mathbb{I}$. According to the property of $\nabla\phi_i(\cdot)$, we have that $\nabla\phi_i(\bm{x}_r^\top\bm{f}_i)\geq0$, with equality holding only when $\bm{x}_r^\top\bm{f}_i=\varepsilon_i^\ast$. Given this, \eqref{Crit} requires that $\bm{x}_r^\top\bm{f}_i<\varepsilon_i^\ast$, and that the vectors $[\bm{x}_r]_\times\bm{x}^{\ast}$ and $[\bm{x}_r]_\times\bm{f}_i$ have the same direction. As shown in Fig. \ref{Critical_points_1}, $[\bm{x}_r]_\times\bm{x}^{\ast}$ points in the opposite direction to the vector $[\bm{x}_r]_\times\bm{f}_0$ for all $\bm{x}_r\in\text{int}(\mathcal{I}_0)$, indicating that no critical points exist in the set $\bm{x}_r\in\mathcal{I}_0$. We next consider the case $\bm{x}_r\in\mathcal{I}_i$, $i\in\mathbb{I}\setminus\{0\}$, where the critical points (if they exist) lie on the plane spanned by $\bm{x}^\ast$ and $\bm{f}_i$, specifically on the side of $\bm{f}_i$ opposite to $\bm{x}^\ast$, as shown in Fig. \ref{Critical_points_2}. This condition ensures that $[\bm{x}_r]_\times\bm{x}^{\ast}$ and $[\bm{x}_r]_\times\bm{f}_i$ point in the same direction. We thus only need to determine whether the magnitude of both sides of \eqref{Crit} are equal. Since $\text{d}_{\mathbb{S}^{2}}(\bm{x}_r,\bm{x}^\ast)<\pi$ and $\text{d}_{\mathbb{S}^{2}}(\bm{x}_r,\bm{f}_i)<\pi$, \eqref{Crit} implies that
\begin{equation}
\label{nabla_U3}
k_a\dfrac{\sin(\text{d}_{\mathbb{S}^{2}}(\bm{x}_r,\bm{x}^\ast))}{\sin(\text{d}_{\mathbb{S}^{2}}(\bm{x}_r,\bm{f}_i))}=k_r\nabla\phi_i(\bm{x}_r^\top\bm{f}_i).
\end{equation}
We note that the left-hand side of \eqref{nabla_U3} varies within a bounded interval $(0,d_i]$ with some $d_i>0$, whereas the right-hand side monotonically increases from $0$ to $\infty$, as $\bm{x}_r^\top\bm{f}_i$ increases from $\varepsilon_i^\ast$ to $\varepsilon_i$. This implies that, for each forbidden zone $\mathcal{P}_i$, $i\in\mathbb{I}\setminus\{0\}$, there certainly exists a single critical point $\bm{x}_r=\bm{c}_i$ that satisfies \eqref{nabla_U3} (equivalent to \eqref{Crit}).  $\hfill \blacksquare$

\bibliographystyle{IEEEtran}
\bibliography{refs}

\vfill

\end{document}